\documentclass[pra,showpacs]{revtex4}

\usepackage{ifpdf}
\ifpdf
\usepackage[dvips]{hyperref}
\else
\usepackage[hypertex]{hyperref}
\fi

\usepackage{graphicx}
\usepackage{natbib}
\usepackage{bm}
\usepackage{amsmath}
\usepackage{amssymb}

\usepackage{color}

\newcommand{\bra}[1]{{\langle{#1}|}}
\newcommand{\ket}[1]{{|{#1}\rangle}}

\newcommand{\ve}[1]{{\overrightarrow{#1}}}
\newcommand{\Tr}{\mathrm{Tr}}
\newcommand{\scal}[2]{\left\langle #1\arrowvert #2\right\rangle}
\newcommand{\vv}[1]{\overrightarrow #1}

\newcommand{\gam}[0]{\overset{\phantom{\rightarrow}}{\gamma}}
\newcommand{\del}[0]{\overset{\phantom{\rightarrow}}{\delta}}
\newcommand{\zg}[0]{\overset{\phantom{\rightarrow}}{0}}

\begin{document}

\title{Composition law for polarizers}

\author{J. Lages}
\email{jose.lages@utinam.cnrs.fr}
\affiliation{Institut UTINAM, UMR CNRS 6213, Equipe de Dynamique des Structures Complexes,
Universit\'e de Franche-Comt\'e, UFR ST, Route de Gray, 25030 Besan\c con Cedex, France}

\author{R. Giust}
\email{remo.giust@univ-fcomte.fr}
\affiliation{FEMTO-ST, UMR CNRS 6174, Universit\'e de Franche-Comt\'e, UFR ST, Route de Gray, 25030 Besan\c con Cedex France}

\author{J.-M. Vigoureux}
\email{jean-marie.vigoureux@univ-fcomte.fr}
\affiliation{Institut UTINAM, UMR CNRS 6213, Equipe de Dynamique des Structures Complexes,
Universit\'e de Franche-Comt\'e, UFR ST, 16 Route de Gray, 25030 Besan\c con Cedex, France}

\begin{abstract}
The polarization process when polarizers act on an optical field is studied.
We give examples for two kinds of polarizers. The first kind presents an anisotropic absorption - as in a polaroid film - and the second one
is based on total reflection at the interface with a birefringent medium.
Using the Stokes vector representation, we determine explicitly the trajectories of the wave light polarization
during the polarization process.
We find that such trajectories are not always geodesics of the Poincar\'e sphere as it is usually thought. Using the analogy between light
polarization and special relativity, we find that the action of successive polarizers on the light wave polarization is equivalent to the action of a single resulting polarizer followed by
a rotation achieved for example by a device with optical activity. We find a composition law for polarizers similar to the composition law for noncollinear velocities in special relativity. We define an angle equivalent to the relativistic Wigner angle which can be used to quantify the quality of two composed polarizers.
\end{abstract}

\maketitle

\section{Introduction}
It is now usual to describe the polarization state of an electromagnetic field by using the notion of Stokes vector and that of the Poincar\'e sphere. This elegant representation of the polarization state is very useful in optics, \textit{e.g.} geometrical phases such as the Pancharatnam phase can be efficiently interpreted on the Poincar\'e sphere \cite{Berry87,Berry90}. Actions of polarizing devices
are described in such a space with simple mathematical operations. As an example, variations of the polarization state when
light passes through an optical active medium is expressed by a simple rotation around the axis connecting the two poles
of the sphere. In the same way, the effect of a birefringent plate corresponds to a rotation around a vector lying in the
equatorial plane by an angle determined by the optical path delay between the ordinary and the extraordinary
axis of the plate.
However, it is curiously interesting to note that the evolution of the polarization due to the action of a polarizer has never been 
studied in details. On the Poincar\'e sphere, such an evolution corresponds to a trajectory intuitively supposed to be a geodesic connecting the polarization states of the field before and after the polarizer (see \textit{e.g.} \cite{Hils99} or \cite{Frins98}). As we will
see, the study of the actions of two different kinds of polarizers (the first one using anisotropic absorbing medium and the second one using total reflection of one component of the electromagnetic
field at the interface with an anisotropic media) will show that such an assumption is not always true and that the trajectories of the polarization state of the field can be more complex and do not necessarily correspond to Poincar\'e sphere geodesics. At the
beginning of the article, we introduce the Jones representation of a polarizer.
The action of a polarizer on the polarization state is then studied by introducing rotation operators and Lorentz boots operators borrowed from special relativity. Since the $SL(2,\mathbb{C})$ group is homomorphic to the Lorentz group (see \textit{e.g.} \cite{Misner73,Halpern68}), a formal equivalence between the special relativity space and the polarization space exists \cite{Kim97,Kim99}. This formal equivalence can then be used to
determine what are the counterpart in the polarization space of, for example, the composition law of velocities, of the aberration phenomenon or of the Wigner angle.

The paper is organized as follows. In section II we define the mathematical formalism used to describe the effect of polarizers on an optical field polarization. The states of polarization of the field are represented on the Poincar\'e
sphere and the action of the optical devices is defined by their associated Jones matrix. It appears that the effect of
polarizers on the polarization are essentially defined by a complex number $A$ which contains informations
about the propagation and the absorption of light in the polarizer. In section III we analyze the evolution of the Stokes vector
$\overrightarrow{s}$ when light goes through different kinds of polarizers. Examples are studied and the trajectories of
given optical states during the polarization process are analyzed. We point out that the projection of such trajectories on the Poincar\'e sphere
are no longer a geodesic of that sphere as it could have been classically expected. In section IV we focus on the evolution of
the optical field intensity and we derive a natural generalized Malus law. Then we examine
the action of two successive polarizers and we show that they are equivalent to one polarizer followed by a rotation exactly as in special relativity the composition of two Lorentz boosts is a Lorentz boost followed by a rotation.
In section V we find the general composition law which permits to determine characteristics of the equivalent
polarizer and the angle of the rotation. Then we discuss the interpretation of such a composition law. In section VI we give the counterpart of the Wigner angle in special relativity for two composed polarizers and we relate such an angle to the quality of the polarizers.

\section{Mathematical description}\label{section_math}

\subsection{Stokes parameters and the Poincar\'e sphere}

The polarization state of light will be described with the use of the Poincar\'e sphere. Our
goal is to describe the evolution of the polarization state during its propagation inside a polarizer. Without loss of
generality we consider that the optical field propagates along the $z$ direction. This field can then be described by the
Jones vector
\begin{equation}\label{wpsi}
    \ket{\widetilde\psi}=
    \left(
        \begin{array}{c}
            E_x \\
            E_y \\
        \end{array}
    \right)_{\{\ket x,\ket y\}}
\end{equation}
where $E_x$ and $E_y$ are the two complex components of the electric field. Instead of the basis $\{\ket x,\ket y\}$
we prefer to work with the basis $\{\ket R,\ket L\}$ related to right- and left-handed circularly polarized states
\begin{equation}
\ket R=\frac{1}{\sqrt{2}}
\left(
\begin{array}{c}
1\\
i
\end{array}
\right)_{\{\ket x,\ket y\}}, \qquad
\ket L=\frac{1}{\sqrt{2}}
\left(
\begin{array}{c}
1\\
-i
\end{array}
\right)_{\{\ket x,\ket y\}}
.
\end{equation}
In this basis the Jones vector $\ket{\widetilde\psi}$ (\ref{wpsi}) is transformed into \cite{Berry96}
\begin{equation}\label{psi}
    \ket{\psi}=\frac{1}{\sqrt{2}}
    \left(
        \begin{array}{cc}
            1&-i\\
            1&\phantom{-}i\\
        \end{array}
    \right)
    \ket{\widetilde\psi}
    =
    \frac{1}{\sqrt{2}}
    \left(
        \begin{array}{c}
            E_x-i\,E_y \\
            E_x+i\,E_y \\
        \end{array}
    \right)_{\{\ket R,\ket L\}}.
\end{equation}

The intensity $I$ and the polarization of the wave light can be characterized by the Stokes parameters \cite{Wolf}
\begin{equation}
\left\{
\begin{array}{l}
s^0=\left\arrowvert E_x\right\arrowvert^2+\left\arrowvert E_y\right\arrowvert^2\equiv I\\
\\
s^1=\left\arrowvert E_x\right\arrowvert^2-\left\arrowvert E_y\right\arrowvert^2\\
\\
s^2=2\,\mathrm{Re}\left(E_x^*E_y\right)\\
\\
s^3=2\,\mathrm{Im}\left(E_x^*E_y\right)
\end{array}
\right..
\end{equation}
If we construct a three dimensional vector $\ve s$ (called hereafter Stokes vector) with the components $s^1$, $s^2$, and $s^3$, it is easy to check that its norm $s=\|\ve s\|=s^0=I$. Thus, for a given intensity, the Stokes vector can be parametrized using spherical coordinates as
\begin{equation}\label{s1}
\ve s=s\left(\sin\theta\cos\phi\vv e^1+\sin\theta\sin\phi\ve
e^2+\cos\theta\ve e^3\right)
\end{equation}
where $\theta\in[0,\pi]$, $\phi\in[0,2\pi[$ and $\{\ve e^1,\ve e^2,\vv e^3\}$ is an $\mathbb{R}^3$ orthonormal basis. The Stokes vector $\ve s$
(\ref{s1}) determines a point located on the Poincar\'e sphere $\mathcal{S}^2$ of radius $s$. Since the norm $s$ gives the intensity of the wave
light, the direction of the vector $\ve s$, \textit{i.e.} the angles $\theta$ and $\phi$, characterizes the polarization. Thus, the polarization state of the light wave corresponds to a unique point on the Poincar\'e sphere $\mathcal{S}^2$. 

\begin{figure}[tb]\centering
    \begin{center}
        \includegraphics[width=0.8\columnwidth]{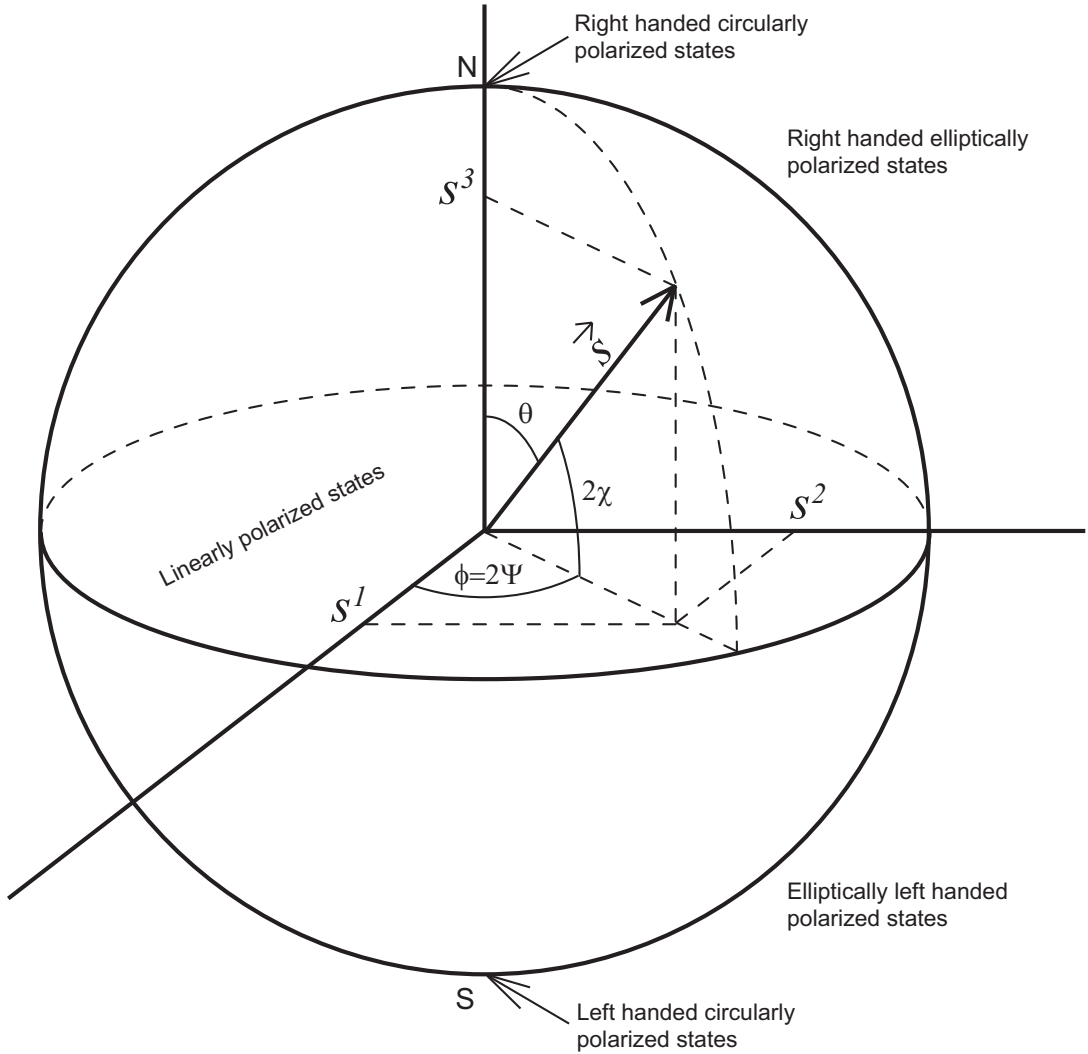}
    \end{center}
    \caption{Representation of the Stokes vector $\vec s$ and the Poincar\'e sphere $\mathcal{S}^2$ }
\end{figure}

In order to follow the evolution of the light wave polarization, \textit{i.e.} the evolution of the Stokes vector $\ve s$, we prefer to work with the projector $\ket\psi\bra\psi$ instead of $\ket\psi$ itself. Indeed, this projector can be easily expressed as a function of $\ve s$ since
\begin{equation}
\ket\psi\bra\psi=\frac{1}{2}\left(
\begin{array}{ll}
s^0+s^3 & s^1-is^2\\
s^1+is^2& s^0-s^3
\end{array}
\right)=\frac{1}{2}\left(s^0\sigma^0+\ve s\cdot\ve \sigma\right)\equiv\rho_{\ve s}.
\end{equation}
Here $\sigma^0$ is the $2\times2$ identity matrix and $\ve \sigma$ is a vector the components of which are the usual Pauli matrices
\begin{equation}
\sigma^1=
\left(
\begin{array}{cc}
0&1\\1&0
\end{array}
\right),\quad
\sigma^2=
\left(
\begin{array}{cc}
0&-i\\i&0
\end{array}
\right),\quad
\sigma^3=
\left(
\begin{array}{cc}
1&0\\0&-1
\end{array}
\right).
\end{equation}

Usually \cite{Wolf}, one choose as the north (south) pole of the Poincar\'e sphere
the circular right-handed (left-handed) polarization state. With this definition, the
angles $\theta$ and $\phi$ (\ref{s1}) are directly related with the
ellipticity angle $\chi$ and the azimuth angle $\Psi$ since
$\theta=\pi/2-2\chi$ and $\phi=2\Psi$ respectively \cite{Wolf}. The
circular right(left)-handed polarization state corresponds to $\theta=0$
($\theta=\pi$).
Right-handed (left-handed) elliptically polarized states correspond to $s^3>0$ ($s^3<0$),
or equivalently $\chi>0$ ($\chi<0$). 
Linear polarization states are represented by
Stokes vectors lying in the equatorial plane of the Poincar\'e Sphere since $\theta=\pi/2$
($\chi=0$). Each linear polarization
states is determined by a given angle $\phi$, or equivalently by a
given azimuth angle $\Psi$. For a linear polarized wave, the angle
$\Psi$ is defined as the angle between the vibration plane and a
laboratory axis $\ve e_x$ (reference axis) orthogonal to the
direction of the propagation. The difference of polarization for two
linear polarized light waves 1 and 2 can be measured, in the
laboratory, as the difference $\Delta\Psi=\Psi_1-\Psi_2$, or
equivalently, on the Poincar\'e sphere, as
$\Delta\phi=\phi_1-\phi_2=2\Delta\Psi$. Thus an angle measured in
the laboratory is half of the corresponding angle in the Poincar\'e
sphere representation. For example, two orthogonal polarizations are
characterized by $\Delta \Psi=\pi/2$ in the laboratory frame and by
$\Delta\phi=\pi$ in the Poincar\'e sphere representation. Although
the rest of the paper does not consider any particular choice of
basis nor any particular choice of polarization, it is useful to
keep in mind these last remarks in order to be able to make at any
time the parallel with known conventions in optics. So, in a general
manner, two states with orthogonal polarizations are represented by
opposite Stokes vectors, and correspond to projectors $\rho_{\ve s}$ and $\rho_{-\ve s}$.

\subsection{Jones matrices for polarizing devices}

\subsubsection{Jones matrix for birefringent systems}

The action of any optical system on the light wave state $\ket{\widetilde\psi}$ (\ref{wpsi}) is defined by
a $2\times 2$ Jones matrix.
For example, the Jones
matrix associated with a birefringent system and acting on $\ket{\widetilde\psi}$ can be written as
\begin{equation}\label{mat1}
\left(
\begin{array}{cc}
e^{i\phi_o}&0\\
0&e^{i\phi_e}
\end{array}
\right)=e^{i(\phi_o+\phi_e)/2}
\left(\begin{array}{cc}
e^{i\phi/2}&0\\
0&e^{-i\phi/2}
\end{array}
\right).
\end{equation}
Here $\phi_o$ and $\phi_e$ are the ordinary and
extraordinary phases associated with the propagation of the optical field polarized along each optical axis.
We have defined the difference between these two phases as $\phi=\phi_o-\phi_e$.
For ordinary and extraordinary axis rotated by an angle $\theta$ in the
($x,y$) laboratory frame, the Jones matrix (\ref{mat1}) becomes
\begin{equation}\label{mat2}
\widetilde B(\theta,\phi_o,\phi_e)=e^{i(\phi_o+\phi_e)/2}
\left(\begin{array}{cc}
\cos\theta&\sin\theta\\
-\sin\theta&\cos\theta
\end{array}
\right)
\left(\begin{array}{cc}
e^{i\phi/2}&0\\
0&e^{-i\phi/2}
\end{array}
\right)
\left(\begin{array}{cc}
\cos\theta&-\sin\theta\\
\sin\theta&\cos\theta
\end{array}
\right).
\end{equation}
The corresponding Jones matrix for a birefringent optical device acting now on the Jones vector $\ket{\psi}$ written in the $\{\ket R,\ket L\}$ basis (\ref{psi}) is then
\begin{equation}\label{mat3}
B(\theta,\phi_o,\phi_e)=\frac12
\left(
        \begin{array}{cc}
            1&-i\\
            1&\phantom{-}i\\
        \end{array}
    \right)
\widetilde B(\theta,\phi_o,\phi_e)
\left(
        \begin{array}{cc}
            1&\phantom{-}1\\
            i&-i\\
        \end{array}
    \right).
\end{equation}
After performing the matrix multiplications, it is possible to recast (\ref{mat3}) in terms of Pauli matrices
\begin{equation}\label{birefrin}
B(\theta,\phi_o,\phi_e)=e^{i(\phi_o+\phi_e)/2}\left(\cos\displaystyle\frac\phi2\,\sigma^0+i\sin\displaystyle\frac\phi2\,\vv p\cdot\overrightarrow{\sigma}\right)
\end{equation}
where $\vv p=\cos2\theta\vv e^1+\sin2\theta\vv e^2$. Using now the mathematical relation $e^{i\alpha\vv{q}\cdot\vv\sigma}=\cos\alpha+i\sin\alpha\vv q\cdot\vv\sigma$ valid for any angle $\alpha$ and any unitary vector $\vv q$, we can rewrite (\ref{birefrin}) in a more compact manner
\begin{equation}\label{birefrin1}
B(\theta,\phi_o,\phi)=e^{i\phi_0}e^{-i\frac\phi2}\,e^{i\frac\phi2\vv{p}\cdot\vv\sigma}.
\end{equation}
As we will see in section \ref{nonabs}, the action
of the matrix $B(\theta,\phi_o,\phi_e)$ on the polarization state $\ket{\psi}$ can be viewed in the Poincar\'e sphere framework as 
the rotation of the Stokes vector $\overrightarrow{s}$ around the vector $\vv p$ by an angle $\phi$.
As no dissipation phenomenon occurs, the norm $s^0$ of the vector $\overrightarrow{s}$ is constant.

\subsubsection{Jones matrix for polarizers}

The Jones
matrix associated with a perfect polarizer acting on $\ket{\widetilde\psi}$ can be written as
\begin{equation}\label{m1}
\left(
\begin{array}{cc}
1&0\\
0&0
\end{array}
\right).
\end{equation}
In that example only the $x$ component of the field is conserved, the orthogonal component is not transmitted. This oversimplified picture corresponds
to the limit case where the attenuation factors in the $x$ and $y$ directions differ from each other by several order of magnitude. The realistic case
correspond then to the matrix
\begin{equation}\label{m2}
\left(
\begin{array}{cc}
e^{-\gamma_1}&0\\
0&e^{-\gamma_2}
\end{array}
\right)
=
e^{-(\gamma_1+\gamma_2)/2}
\left(
\begin{array}{cc}
e^{\gamma/2}&0\\
0&e^{-\gamma/2}
\end{array}
\right)
\end{equation}
where $e^{-\gamma_1}$ and $e^{-\gamma_2}$ are the attenuation factors in the $x$ and $y$ directions respectively. We have defined the difference
between the two attenuation terms as $\gamma=\gamma_2-\gamma_1$. Again, if the polarizer axis is rotated about an angle $\theta$ in the
($x,y$) laboratory frame, the Jones matrix (\ref{m2}) becomes
\begin{equation}\label{m3}
\widetilde P(\theta,\gamma_1,\gamma_2)=e^{-(\gamma_1+\gamma_2)/2}
\left(\begin{array}{cc}
\cos\theta&\sin\theta\\
-\sin\theta&\cos\theta
\end{array}
\right)
\left(
\begin{array}{cc}
e^{\gamma}&0\\
0&e^{-\gamma}
\end{array}
\right)\left(\begin{array}{cc}
\cos\theta&-\sin\theta\\
\sin\theta&\cos\theta
\end{array}
\right).
\end{equation}
Following the same steps as in the previous section for birefringent systems, the Jones matrix for a polarizer acting on $\ket{\psi}$
in the $\{\ket R,\ket L\}$ basis (\ref{psi}) is
\begin{equation}\label{m4}
P(\theta,\gamma_1,\gamma)=e^{-\gamma_1}e^{-\frac\gamma2}
\,e^{\frac\gamma2\vv{p}\cdot\vv\sigma}
\end{equation}
where the vector $\vv p$ encodes again the information about the direction of the polarizer in the laboratory frame. In the Poincar\'e sphere
framework, as the attenuation difference $\gamma$ increases, the action of the operator (\ref{m4}) on $\ket\psi$ can be seen as the progressive collapse of the Stokes vector $\vv s$ on the polarizer vector $\vv p$ direction (see section \ref{sabs}).

We now use the derived expressions (\ref{birefrin1}) and (\ref{m4}) to examine more precisely the case of a polarizer based on
anisotropic absorption (such as a polaroid film), and the case of a polarizer based on a total reflection at the interface of
an anisotropic crystal.

The case of an anisotropic absorbing polarizer can be modeled by a system where the phases $\phi_o$ and $\phi_e$
in orthogonal directions are complex,
\begin{equation}\label{eq_abs}
\begin{array}{ccc}
    \phi_o&=&(k+i\alpha_o)z,\\
    \phi_e&=&(k+i\alpha_e)z.
\end{array}
\end{equation}
The parameter $k$ is the wavevector of the optical field in the film and $\alpha_o$ and $\alpha_e$ are the two absorption
coefficients (for the amplitude) in two orthogonal directions of polarization. The system acts as a linear polarizer if
$\alpha_o\ne\alpha_e$. In equations (\ref{eq_abs}), the parameter $z$ is the propagation coordinate, \textit{i.e.} $z=0$ at the entrance of the
polarizer and, \textit{e.g.} $z=L$ at its end. Hence, using (\ref{birefrin1}), the operator acting on $\ket\psi$ associated to such a polarizer
is
\begin{equation}\label{pp1}
P_1(\theta,\phi_o,\gamma)=e^{ikz}e^{-\alpha_oz}e^{-\frac\gamma2}
\,e^{\frac\gamma2\vv{p}\cdot\vv\sigma}
\end{equation}
where $\gamma=(\alpha_e-\alpha_o)z$.
Up to a global phase term and a global attenuation term which do not affect the light wave polarization, we retrieved, as expected, the expression of
the Jones matrix for a polarizer (\ref{m4}).

The other case corresponds to a polarizer based on total reflection. We consider a polarized light hitting the face of an anisotropic crystal in such a manner that only one component of the polarized light is transmitted, the other component being evanescent after the interface.
This kind of polarizer can be modeled by an optical device with the phase $\phi_o$ and $\phi_e$
defined as
\begin{equation}
\begin{array}{ccc}
    \phi_o&=&k\,z,\\
    \phi_e&=&i\,\mu\,z
\end{array}
\end{equation}
where $k$ and $\mu$ are the $z$ real components of the wavevector of the linear polarization which are transmitted and totally
reflected respectively. Here, the parameter $z$ is the propagation coordinate along the transmitted light wave component ($z=0$ at the interface). Again, using (\ref{birefrin1}), the Jones operator acting on the wave light $\ket\psi$ and associated to such polarizer is
\begin{equation}\label{pp2}
P_2(\theta,\delta,\gamma)=e^{ikz}e^{-\frac{\gamma+i\delta}2}
\,e^{\frac{\gamma+i\delta}2\vv{p}\cdot\vv\sigma}
\end{equation}
where $\gamma=\mu z$ and $\delta=kz$.

Comparing the just derived expressions (\ref{birefrin1}), (\ref{m4}), (\ref{pp1}) and (\ref{pp2}) of Jones operators (Jones matrices) for polarizing devices, we remark that all of them are written in the following manner
\begin{equation}\label{fin}
P(\vv p,\phi_p,A)=e^{i\phi_p}\,e^{-\frac A 2}\,e^{\frac{A}{2}\,\overrightarrow{p}\cdot\overrightarrow{\sigma}}
\end{equation}
where $A$ is a complex differential attenuation term, $\phi_p$ an overall phase/attenuation term, and $\vv p$
the polarization vector associated with the axis of the polarizer.
In the following we will drop out the overall factor $e^{i\phi_p}$ from the expression of the Jones operator (\ref{fin}) since it does not
affect the evolution of the wave light polarization. We have just to keep in mind that in the case of linear polarizer such as those defined in
(\ref{m4}) and (\ref{pp1}), the overall factor $e^{i\phi_p}$ contains an attenuation part which globally reduces the intensity of the wave light.
From now on, it will be useful to split $A$ in its
real and imaginary parts, $A=\gamma+i\,\delta$. Note that $A$, and consequently $\gamma$ and $\delta$, are implicit linear functions of the $z$ propagation coordinate.

Also, although equation (\ref{fin}) has been derived from the analysis of linear polarizers 
(vector $\vv p$ lying in the equatorial plane), our model is completely general.
It is possible to use a more general rotation than (\ref{m3}) bringing the polarizer
characteristic vector $\vv p$ outside the equatorial plane.  For example, if the
vector $\vv p$ points toward the Poincar\'e sphere North (or South) pole, our model (\ref{fin})
describes circular polarizers or, if $A=i\delta$, media with optical activity.

\section{Light wave polarization viewed as a Stokes 4-vector transformation}

According to the previous section, the action of a polarizing device on a light wave can be characterized by the following operator
\begin{equation}\label{op}
    P_{\ve p,\gamma,\delta}(z)=e^{-\frac 1 2\left(\gamma+i\delta\right)}e^{\frac 1 2\left(\gamma+i\delta\right) \ve p\cdot\vv\sigma},
\end{equation}
where $\gamma\left(z\right)$ and $\delta\left(z\right)$ are real positive functions of the transformation parameter
$z\in\mathbb{R}^+$. Usually, $z$ is the length penetration of light into the device and the functions $\gamma(z)$ and $\delta(z)$
encode the absorption and the propagation of light inside the device respectively. We restrict our work to homogeneous media for
which these functions are linear in $z$, \textit{i.e.} $\gamma(z)=\alpha z$ and $\delta(z)=\beta z$ with
$\alpha,\beta\in\mathbb{R}$. The vector $\ve p$ is a normalized polarization vector ($\left\|\ve p\right\|=p=1$) associated with
the polarizing device. The vector $\ve p$ determines a point on the Poincar\'e sphere which corresponds to a pure state
$\rho_{\ve p}$ of polarization $\ve p$.

Using the projector $\rho_{\ve s}=\ket{\psi}\bra{\psi}$ we are able to define the Stokes 4-vector $\mathbf{s}=(s^0,\ve s)$ with the
following components
\begin{equation}\label{stokes1}
s^\mu=\Tr\left(\sigma^\mu\rho_{\ve s}\right), \quad \mu\in\{0,1,2,3\}.
\end{equation}
It is interesting to note that, introducing the Minkowskian metric $\eta_{\mu\nu}=\mathrm{diag}\{-1,1,1,1\}$, the Stokes 4-vector can be
considered as a \textit{light-like} 4-vector since $\eta_{\mu\nu}s^\mu s^\nu=0$.
The spatial components of the Stokes 4-vector,
$\{s^i\}_{i=1\dots3}$, are the components of the three dimensional Stokes vector $\ve s$ defined in the previous section.
The \textit{time} component of the Stokes
4-vector $s^0$ is the norm of the Stokes vector $\ve s$. Consequently this last component also corresponds to the light wave intensity
$s^0=\scal{\psi}{\psi}=I$.

So, according to the operator (\ref{op}), a polarization state $\rho_{\ve s(0)}$ is continuously transformed into another polarization state $\rho_{\ve s(z)}$ by the following operation
\begin{equation}\label{rhoevol}
    \rho_{\ve s(z)}=P_{\ve p,\gamma,\delta}(z)\rho_{\ve s(0)}P^\dagger_{\ve p,\gamma,\delta}(z), \quad z\in\mathbb{R}^+.
\end{equation}
Now, using (\ref{stokes1}) and (\ref{rhoevol}) we are able to define the evolution of the Stokes 4-vector $\mathbf{s}(z)$ by
\begin{equation}\label{stcp}
    \begin{array}{lcl}
        s^\mu(z)&=&\Tr\left(\sigma^\mu\rho_{\ve s(z)}\right)\\
        &=&\Tr\left(
        P^\dagger_{\ve p,\gamma,\delta}(z)\sigma^\mu P^{\phantom{\dagger}}_{\vv
        p,\gamma,\delta}(z)\rho_{\ve s(0)}\right), \quad
        \mu\in\{0,1,2,3\}
    \end{array}
\end{equation}
where we recall that $s^0(z)=s(z)=\left\|\ve s(z)\right\|=I(z)$ is the intensity of the light wave, and $\{s^i(z)\}_{i\in\{1,2,3\}}$
are the spatial components of the Stokes vector $\vv s(z)$.
After some algebraic calculus, the components $s^\mu(z)$ of the Stokes 4-vector (\ref{stcp})  can be expressed as
\begin{widetext}
    \begin{equation}\label{s4v}
    s^\mu(z)=
    \left\{
        \begin{array}{lcl}
            s^0(z)&=&e^{-\gamma(z)}\left(\,s^0(0)\cosh\gamma(z)+\sinh\gamma(z)\ve s(0)\cdot\ve p\,\right)\\
            \\
            \vv{s}(z)&=&e^{-\gamma(z)}\big(\,\ve s(0)\cos\delta(z)+\ve s(0)\wedge\vv p\sin\delta(z)
            +\left(1-\cos\delta(z)\right)\left(\ve p\cdot\vv
            s(0)\right)\ve p\\
            &&\phantom{e^{-\gamma(z)}}
            +s^0(0)\vv p\sinh\gamma(z)+\left(\cosh\gamma(z)-1\right)\left(\ve p\cdot\vv
            s(0)\right)\ve p\,\big)
        \end{array}
    \right.
    \end{equation}
\end{widetext}
For the sake of clarity we now consider the two particular cases $\gamma=0\,,\delta\neq 0$ and $\gamma\neq 0\,,\delta=0$ before
considering the more general case of polarizers with $\gamma\neq0$ and $\delta\neq0$.

\subsection{Non absorbing polarizing device : $\gamma=0$, $\delta\neq 0$}\label{nonabs}
\begin{figure}[tb]\centering
    \begin{center}
        \includegraphics[width=0.8\columnwidth]{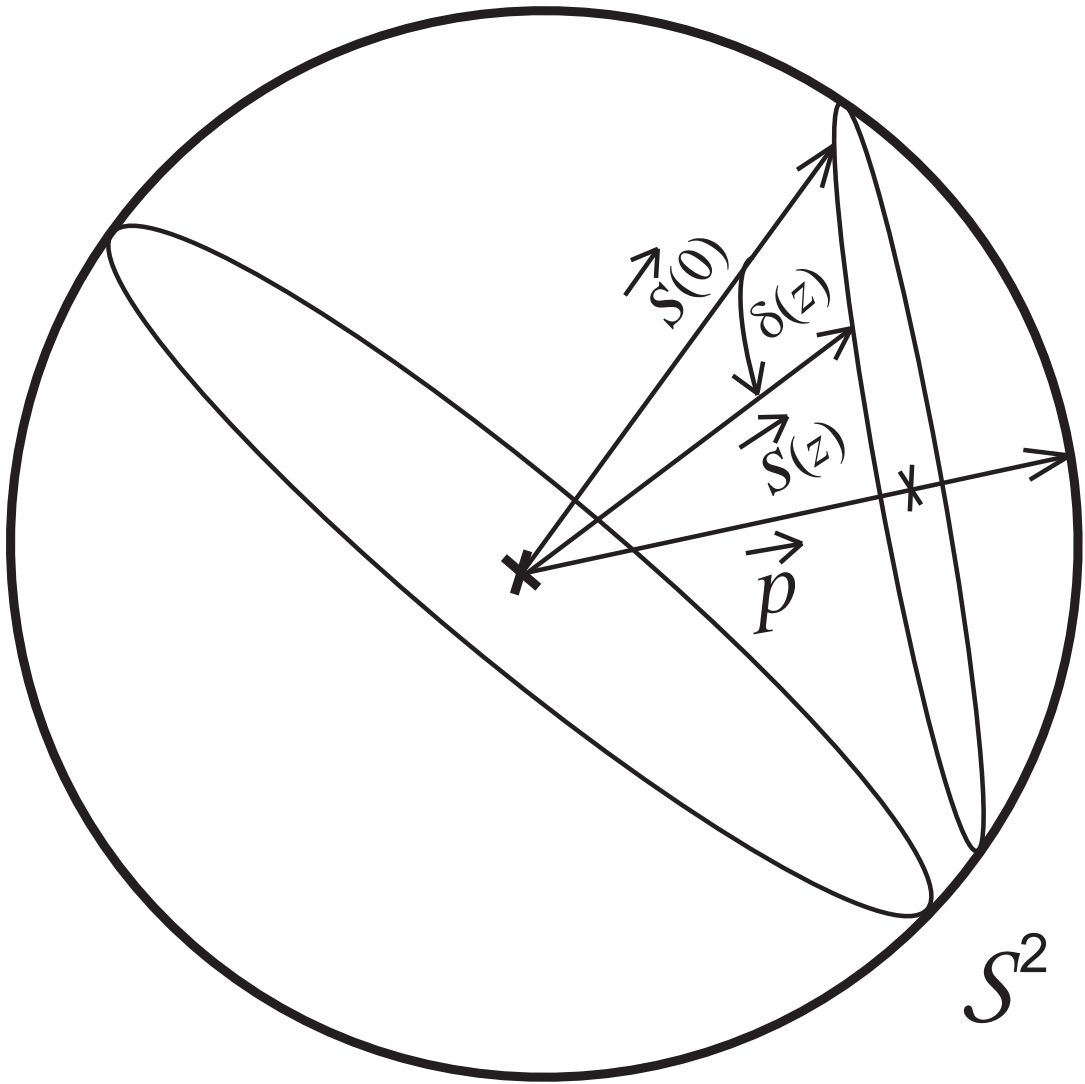}
    \end{center}
    \caption{\label{pna}In the case of a non absorbing polarizing device, the Stokes vector $\vec s$ is rotated around the polarization vector $\vec p$.}
\end{figure}

The unitary case $\gamma=0$, $\delta\neq0$, $\forall z\in\mathbb{R}$ corresponds to non absorbing birefringent devices or to
media with optical activity. In such a case, the transformation (\ref{s4v}) of the Stokes 4-vector $\mathbf{s}$ becomes
\begin{equation}\label{s4vna}
    s^\mu(z)=
    \left\{
    \begin{array}{lcl}
        s^0(z)&=&s^0(0)\equiv1\\
        &&\\
        \vv{s}(z)&=&\ve s(0)\cos\delta(z)\\
        &+&\ve s(0)\wedge\vv p\sin\delta(z)\\
        &+&\left(1-\cos\delta(z)\right)\left(\ve p\cdot\vv
        s(0)\right)\ve p.
    \end{array}
    \right.
\end{equation}
The first equation in (\ref{s4vna}) verifies that $s(z)=\left\|\ve s(z)\right\|=s^0(0)\equiv1$ is invariant under the action of
the non absorbing polarizing device. So, the light wave intensity $I$ is conserved and for convenience we have set $I\equiv1$. The extremity of the vector $\ve s(z)$ still
lies on the Poincar\'e unit sphere after the transformation. The second equation in (\ref{s4vna}) is the Rodrigues Formula for
the rotation. This equation implies that the Stokes vector $\ve s(0)$ has been rotated around the vector $\ve p$ by an angle
$\delta(z)$ (Fig. \ref{pna}).
Without any loss of generality, we can always choose an $\mathbb{R}^3$ orthonormal basis $\{\ve e^i\}_{i\in\{1,2,3\}}$ such as $\ve
p=\ve e^1$. Then the Stokes 4-vector transformation (\ref{s4vna}) can be written in the following simple form
\begin{equation}
    \mathbf{s}(z)= \left(
    \begin{array}{cccc}
        1&0&0&0\\
        0&1&0&0\\
        0&0&\cos\delta(z)&\sin\delta(z)\\
        0&0&-\sin\delta(z)&\cos\delta(z)
    \end{array}
    \right)
    \mathbf{s}(0)
\end{equation}
where we recognize the rotation matrix around the $\ve e^1$ axis by an angle $\delta(z)$.

The transformation (\ref{s4vna}) of the light wave polarization presented in Fig. \ref{pna} is typical of birefringent devices. For example a $\lambda/4$ plate can be used to transform a linear polarized light wave into a circular polarized one, \textit{i.e.} 
to transform a Stokes vector lying in the equatorial plane
into a Stokes vector pointing toward a pole of the Poincar\'e sphere.

\subsection{Absorbing polarizer : $\gamma\neq0$, $\delta=0$}\label{sabs}
\begin{figure}[tb]
    \begin{center}
        \includegraphics[width=0.8\columnwidth]{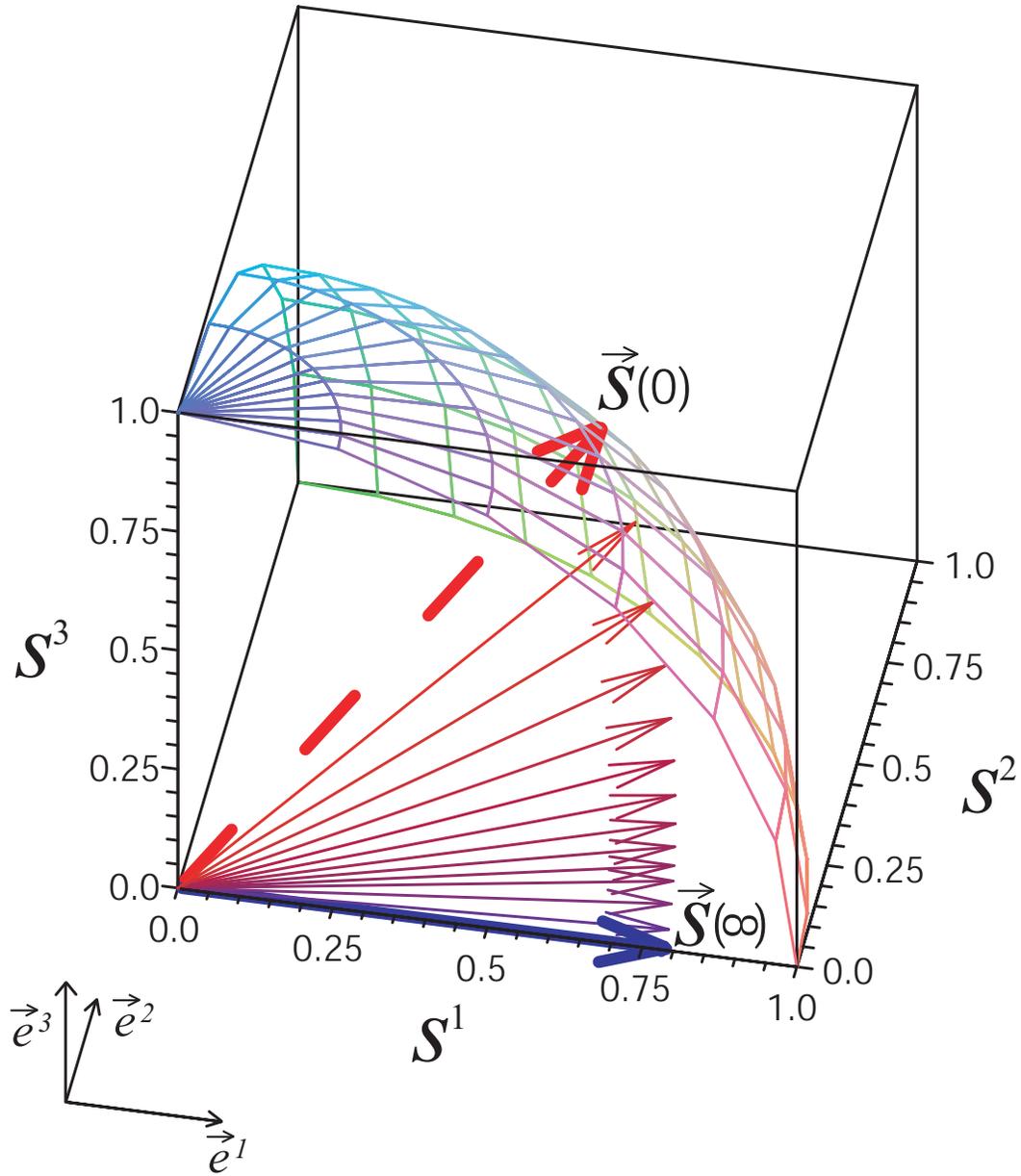}
    \end{center}
    \caption{\label{figbbb}(Color online) Transformation (\ref{abs}) of the spatial components $\vec s(z)$ of the Stokes 4-vector
    $\mathbf{s}(z)$. The polarization vector is $\vec p=\vec e^{\,1}$.  The incoming light wave has an initial polarization
    $\vec s(0)=1/\sqrt{3}\left(\vec e^{\,1}+\vec e^{\,2}+\vec e^{\,3}\right)$ represented by the thick red (light gray)
    dashed vector.
    The Stokes vector for
    $\gamma\rightarrow\infty$ is
    alignated along the polarization vector $\vec p=\vec e^1$ and is represented by the thick blue (dark gray) vector.
    The other vectors are drawn to show the intermediate positions of the Stokes vector $\vec{s}$
    as the parameter $\gamma$ (or equivalently the parameter
    $z$) increases from $0$ to $\infty$.}
\end{figure}
We consider now the non-unitary case $\gamma\neq0$, $\delta=0$, $\forall z\in\mathbb{R}^+$. In such a case, the transformation
(\ref{s4v}) of the Stokes 4-vector $\mathbf{s}$ becomes
\begin{widetext}
    \begin{equation}\label{abs1}
        s^\mu(z)=
        \left\{
            \begin{array}{lcl}
                s^0(z)&=&e^{-\gamma(z)} \left(
                s^0(z)\cosh\gamma(z)+\sinh\gamma(z)\ve s(0)\cdot\ve p\right)\\
                &&\\
                \vv{s}(z)&=&e^{-\gamma(z)} \left(\ve s(0) +s^0(0)\vv
                p\sinh\gamma(z)+\left(\cosh\gamma(z)-1\right)\left(\ve p\cdot\vv
                s(0)\right)\ve p\right)
            \end{array}
        \right.
    \end{equation}
\end{widetext}
Naturally, the $s^0$ component giving the intensity of the light wave (first equation in (\ref{abs1})) is a decreasing function
of the parameter $\gamma$. As $\gamma\rightarrow\infty$, the Stokes vector is brought along the direction of the $\ve p$ vector
since
\begin{equation}
    \ve s(z)\overset{\gamma\rightarrow+\infty}{\propto}\ve p.
\end{equation}
For a finite value of $\gamma$, the Stokes vector is not completely brought along the direction of the vector $\ve p$, the
polarization of the light wave is then only partial. From now on, devices which transform the light wave polarization according to
(\ref{abs1}) will be designated as non perfect polarizers (finite $\gamma$) or perfect polarizer ($\gamma\rightarrow\infty$).

\begin{figure}[ftb]
    \begin{center}
        \includegraphics[width=0.8\columnwidth]{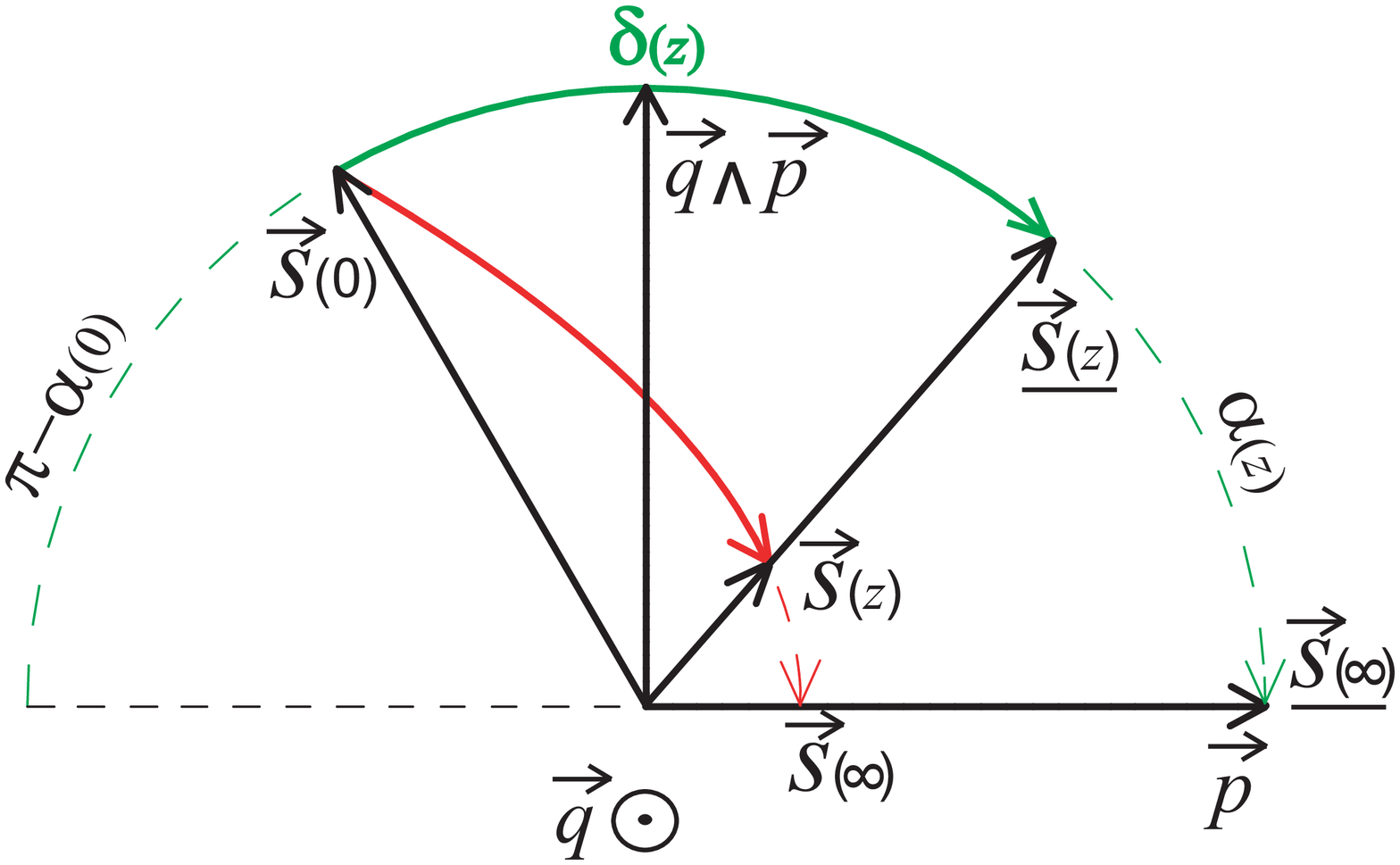}
    \end{center}
    \caption{\label{stabs}(Color online) Representation of typical evolutions of the Stokes vector $\vec s(z)$ and of the normalized
    Stokes vector $\underline{\vec s}(z)$ according to the transformations (\ref{abs1}) and  (\ref{absnorm}) respectively.}
\end{figure}

\subsubsection{Four dimensional representation}
Again, without any loss of generality, we can choose $\vv p=\ve e^1$. Then the transformation (\ref{abs1}) reads
\begin{equation}\label{abs}
    \mathbf{s}(z)=e^{-\gamma(z)} \left(
    \begin{array}{cccc}
        \cosh\gamma(z)&\sinh\gamma(z)&0&0\\
        \sinh\gamma(z)&\cosh\gamma(z)&0&0\\
        0&0&1&0\\
        0&0&0&1
    \end{array}
    \right)\mathbf{s}(0).
\end{equation}
We recognize here a Lorentz boost in the $\ve p=\ve e^1$ direction, the absorption term $\gamma$ is then similar to the
\textit{rapidity} $\Phi=\arg\tanh v/c$ in special relativity. So, up to an attenuation factor $e^{-\gamma}$,
the transformation (\ref{abs1}) is a Lorentz boost in the direction
of the polarizer vector $\ve p$. The Lorentz boost mixes only the intensity component $s^0(0)$ and the component along the
polarizer vector, \textit{i.e.} the projection $\ve s(0).\ve p$. The other components are left unchanged by the boost. Associated
with the global attenuation $e^{-\gamma(z)}$, the boost attenuates the components of $\vv s$ which are orthogonal to the
polarizer vector $\ve p$: $\ve s$ is progressively brought in the direction of the polarization vector $\ve p$ (Fig.
\ref{figbbb}).

Note that this case exactly corresponds to using a polaroid film with $\gamma=\left(\alpha_e-\alpha_o\right)z$.

\subsubsection{The Poincar\'e sphere representation}
The evolution of the light wave polarization can also be followed on the surface of the Poincar\'e sphere by studying the
evolution of the normalized Stokes vector $\underline{\vv{s}}(z)=\vv{s}(z)/\left\|\ve s(z)\right\|$
\begin{widetext}
    \begin{equation}\label{absnorm}
        \underline{s}^\mu(z)=\left\{
        \begin{array}{lcl}
            \underline{s}^0(z)&=&\underline{s}^0(0)=s^0(0)\equiv1\\
            &&\\
            \underline{\vv{s}}(z)&=&\displaystyle \frac{\ve s(0) +\vv
            p\sinh\gamma(z)+\left(\cosh\gamma(z)-1\right)\left(\ve p\cdot\vv
            s(0)\right)\ve p}{s^0(0)\cosh\gamma(z)+\sinh\gamma(z)\ve s(0)\cdot\ve p}
        \end{array}
        \right.
    \end{equation}
\end{widetext}
We remark easily that throughout the continuous transformation the Stokes vector $\ve s(z)$ is always orthogonal to the unit
vector
\begin{equation}
    \vec q=\frac{1}{s^0(0)\sin\alpha(0)}\ve p\wedge \ve s(0)
\end{equation}
 where $\alpha(z)=\widehat{\left(\ve p,\ve
s(z)\right)}$.
Indeed, for any parameter $z$, we can write
\begin{equation}
    \ve s(0) \wedge \ve s(z)\propto\ve q.
\end{equation}
The set of the Stokes vectors $\ve s(z)$ with $z \in\mathbb{R}^+$ defines a plane orthogonal to the vector $\ve q$ and containing the
center of the Poincar\'e sphere. This implies that the normalized Stokes vector $\underline{\ve s}(z)$ describes an arc of a
great circle of the Poincar\'e sphere $\mathcal{S}^2$ (Fig. \ref{stabs}). The projection of the Stokes vector $\vv s(z)$ describes a geodesic of the $\mathcal{S}^2$ unit Poincar\'e sphere. Therefore, the normalized Stokes vector $\underline{\ve
s}(z)$ is rotated around the axis $\vec q$ by an angle $\delta(z)=\alpha(0)-\alpha(z)$ (see Fig. \ref{stabs}). Using the Rodrigues formula, we can rewrite the transformation (\ref{absnorm}) as a simple
rotation
\begin{equation}\label{absrot}
    \underline{s}^\mu(z)=\left\{
    \begin{array}{lcl}
        \underline{s}^0(z)&=&\underline{s}^0(0)\equiv1\\
        &&\\
        \underline{\vv{s}}(z)&=&\ve s(0) \cos\delta(z)+\ve s(0)\wedge\vv
        q\sin\delta(z)
    \end{array}
    \right.
\end{equation}
with
\begin{equation}
\sin\delta(z)=\sin\alpha(0)\cos\alpha(z)-\sin\alpha(z)\cos\alpha(0)
\end{equation}
and
\begin{equation}
\cos\delta(z)=\cos\alpha(0)\cos\alpha(z)+\sin\alpha(z)\sin\alpha(0)
\end{equation}
where
\begin{equation}\label{coscoscos}
\cos\alpha(z)=\frac{\tanh\gamma(z)+\cos\alpha(0)}{1+\tanh\gamma(z)\cos\alpha(0)}
\end{equation}
and
\begin{equation}
\sin\alpha(z)=\frac{\sin\alpha(0)}{\cosh\gamma(z)+\sinh\gamma(z)\cos\alpha(0)}.
\end{equation}

\begin{figure}[t]
    \begin{center}
    \includegraphics[width=0.8\columnwidth]{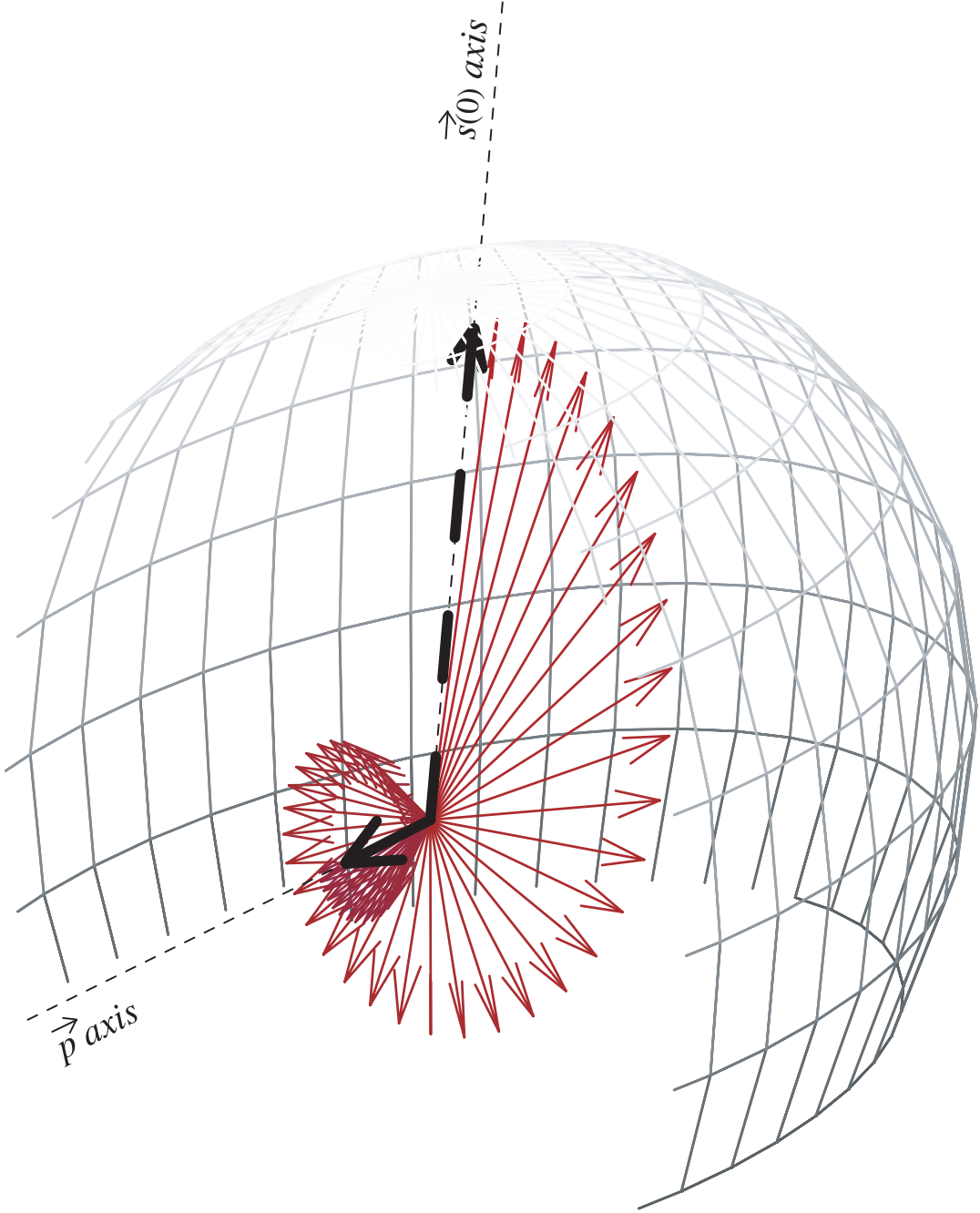}
    \end{center}
    \caption{\label{figgen}(Color online)
    Typical representation of the Stokes vector $\vec s(z)$ evolution according to the general
    transformation (\ref{gen}) or (\ref{s4v}).
    The two dashed lines give the direction of the initial Stokes vector $\vec s(0)$
    and the
    direction of the polarization vector $\vec p$ respectively.
    The initial Stokes vector $\vec s(0)$ is represented by the thick black dashed vector.
    The Stokes vector for $\gamma\rightarrow\infty$ is represented by the thick black vector
    oriented along the direction of the polarization
    vector $\vec p$.
    The other vectors are the successive Stokes vectors $\vec s$ as the parameter $\gamma$ (or equivalently the parameter
    $z$) increases from $0$ to $\infty$.} 
\end{figure}

\subsection{General polarizer : $\gamma\neq0$, $\delta\neq0$}\label{sec:gen}

Choosing again $\ve p=\ve e^1$, the transformation (\ref{s4v}) for a general polarizer can be written in the following matrix
form
\begin{eqnarray}\notag
    \mathbf{s}(z)&=&e^{-\gamma(z)}
    \left(
    \begin{array}{cccc}
        \cosh\gamma(z)&\sinh\gamma(z)&0&0\\
        \sinh\gamma(z)&\cosh\gamma(z)&0&0\\
        0&0&1&0\\
        0&0&0&1
    \end{array}
    \right)\\\label{gen}
    &&\phantom{e^{-\gamma(z)}}\cdot
    \left(
    \begin{array}{cccc}
        1&0&0&0\\
        0&1&0&0\\
        0&0&\cos\delta(z)&\sin\delta(z)\\
        0&0&-\sin\delta(z)&\cos\delta(z)
    \end{array}
    \right)
    \mathbf{s}(0).
\end{eqnarray}
Equation (\ref{gen}) shows clearly that the action of a polarizer on a light wave can be seen as three successive operations: a
rotation of the Stokes vector  $\ve s(0)$ by an angle $\delta(z)$ around the polarizer vector $\ve p$, a Lorentz boost in the
direction of the polarizer vector $\ve p$, and the global attenuation $e^{-\gamma(z)}$. Each of these three operations evidently
commutes with the others. Thus, the operator (\ref{op}) associated with a general polarizing device can be rewritten as
\begin{equation}\label{op1}
    P_{\ve p,\gamma,\delta}(z)=e^{-\frac 1 2\gamma\left(z\right)}e^{-\frac i 2\delta\left(z\right)}
    R_{\ve p,\delta}(z)B_{\ve p,\gamma}(z)
\end{equation}
where in the $SL(2,\mathbb{C})$ group representation
\begin{equation}\label{sl2cb}
    B_{\vv p,\gamma}(z)=e^{\frac{\gamma(z)}{2} \vv p \cdot \vv \sigma}
\end{equation}
is the Lorentz boost operator in the direction $\vv p$ with the \textit{rapidity} $\gamma$ and
\begin{equation}\label{sl2cr}
    R_{\vv q,\delta}(z)=e^{i\frac{\delta(z)}{2} \vv q \cdot \vv \sigma}
\end{equation}
is the rotation operator with the rotation vector $\ve q$ and the angle of rotation $\delta$. Fig. \ref{figgen} gives a typical
illustration of the Stokes vector evolution for a general polarizer (\ref{s4v}).

Note that this case corresponds to total reflection based polarizers (\ref{pp2}) with $\gamma=\mu z$ and $\delta=-kz$.

As a conclusion of this section, we can state that during the polarization process of a wave light, the evolution of the Stokes vector, \textit{i.e.}
the evolution of the polarization, does not necessarily describe a geodesic on the Poincar\'e sphere.
A geodesic, \textit{i.e.} here a part of a great circle of $\mathcal{S}^2$, is found only in the case of polarizing device such as a $\lambda/4$ plate or in the case of non-unitary polarizer with $\delta=0$ such as \textit{e.g.} a polaroid film. In this last case, it is the projection of the Stokes vector on the Poincar\'e sphere which describe a geodesic. All these statements on the effective trajectory of the light wave polarization are worth to consider when geometric phase have to be calculated \cite{lgv2008}.

\section{Generalized Malus law, degree of polarization}\label{maluschap}
\begin{figure*}[tb]
    \begin{center}
        \includegraphics[width=0.7\textwidth]{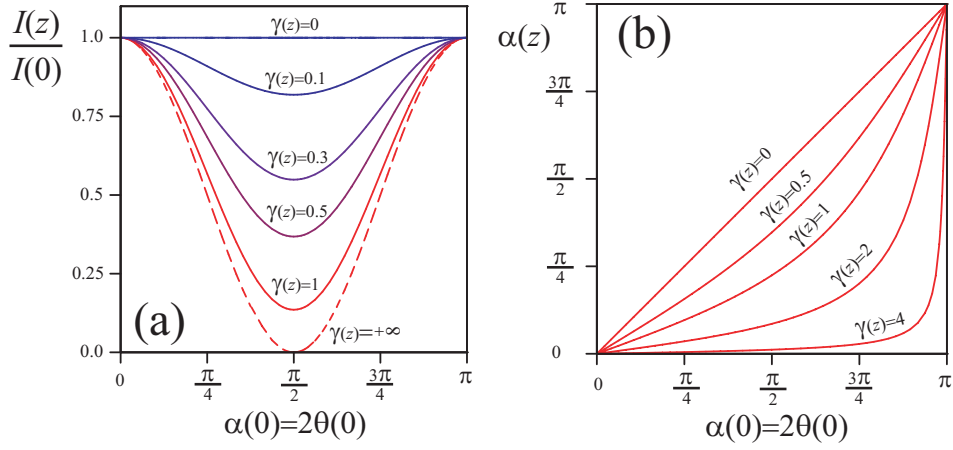}
    \end{center}
    \caption{\label{figmalus}(Color online) (a) Generalized Malus law: representation of the ratio $I(z)/I(0)$ as a function of
    $\alpha(0)=2\theta(0)$ (angle between the polarizer vector $\vec p$ and the incoming wave polarization $\vec s(0)$) for different values
    of the absorption term $\gamma(z)$. (b) Angle $\alpha(z)$ between the polarizer vector $\vec p$ and the wave polarization $\vec s(z)$
    as a function of $\alpha(0)$ for different values of the absorption term $\gamma(z)$. }
\end{figure*}

Now, let us look at the intensity component $s^0(z)$ of the Stokes 4-vector. We define $\alpha(z)=2\theta(z)$ as the oriented
angle $\widehat{\left(\ve p,\ve s(z)\right)}$ between the polarizer vector $\ve p$ and the Stokes vector $\ve s(z)$ for the
parameter $z$. Then, $\theta(z)$ is the angle, \textit{in the laboratory frame}, between the axis of the polarizer and the ordinary
axis of the light wave at position $z$. The intensity of the light wave, given by the first equation in (\ref{s4v}), can be
rewritten as
\begin{equation}\label{malus1}
    I(z)=I(0)\left(\cos^2\theta(0)+\sin^2\theta(0)e^{-2\gamma(z)}\right)
\end{equation}
which is a generalization of the Malus law for non perfect polarizers. Fig. \ref{figmalus}a shows the Malus law for different value
of the absorption term $\gamma(z)$ from $0$ to $+\infty$.
Let us define $I_{\parallel}(z)=I_{\ve p}(z)$ as the intensity of the light wave in the $\rho_{\ve p}$ polarizer state and
$I_{\perp}(z)=I_{-\ve p}(z)$ as the intensity in the corresponding orthogonal state $\rho_{-\ve p}$. After some calculus we
obtain
\begin{equation}
    I_{\parallel}(z)=I_{\ve p}(z)=\Tr{\left(\rho_{\ve s(z)}\rho_{\ve p}\right)}=I(0)\cos^2\theta(0)
\end{equation}
and
\begin{equation}
    I_{\perp}(z)=I_{-\ve p}(z)=\Tr{\left(\rho_{\ve s(z)}\rho_{-\ve p}\right)}=I(0)\sin^2\theta(0)e^{-2\gamma(z)}.
\end{equation}
Now, if the polarizer is placed in an unpolarized beam the transmitted intensity at the point $z$ is
\begin{equation}\label{malusunpo}
    \left\langle I(z)\right\rangle_{\theta(0)}=\left\langle I_{\parallel}(z)\right\rangle_{\theta(0)}+\left\langle I_{\perp}(z)\right\rangle_{\theta(0)}
    =\frac{I(0)}{2}\left(1+e^{-2\gamma(z)}\right)
\end{equation}
where the brackets $\left\langle \cdots\right\rangle_{\theta(0)}$ denote the average over the initial angles $\theta(0)$. From (\ref{malusunpo}), we remark that for a non perfect polarizer ($\gamma$ finite) more than the half of the incoming intensity is transmitted. The degree of polarization $\eta(z)$ of the non perfect polarizer is given by
\begin{equation}\label{dop}
    \eta(z)=\frac{\left\langle I_{\parallel}(z)\right\rangle_{\theta(0)}-\left\langle I_{\perp}(z)\right\rangle_{\theta(0)}}{\left\langle I_{\parallel}(z)\right\rangle_{\theta(0)}+\left\langle I_{\perp}(z)\right\rangle_{\theta(0)}}=\tanh\gamma(z)
\end{equation}
and the extinction ratio $\xi(z)$ by
\begin{equation}\label{ex}
    \xi(z)=\frac{\left\langle I_{\perp}(z)\right\rangle_{\theta(0)}}{\left\langle I_{\parallel}(z)\right\rangle_{\theta(0)}}=e^{-2\gamma(z)}.
\end{equation}
As $\gamma(z)$ varies from $0$ to $+\infty$, $\eta(z)$ and $\xi(z)$ vary from $0$ to $1$ and from $1$ to $0$ respectively.
Although $\eta(z)$ and $\xi(z)$ measure already the efficiency of the polarizer, we can always ask \textit{how far} is the
polarization vector $\ve s(z)$ from the polarization vector $\ve p$ associated with the polarizer. For that purpose
we focus on the angle $\alpha(z)$ between the two vectors. Combining the two equations in (\ref{s4v}), or equivalently using (\ref{dop}) and (\ref{coscoscos}), we obtain the following
relation
\begin{equation}\label{exp1}
    \cos\alpha(z)=\displaystyle\frac{\eta(z)+\cos\alpha(0)}{1+\eta(z)\cos\alpha(0)}.
\end{equation}
from which we are able to extract the angle $\alpha(z)$ as
\begin{equation}\label{exp2}
    \alpha(z)=\arccos\left(\tanh\left(\gamma(z)+\arg\tanh\left(\cos\alpha(0)\right)\right)\right).
\end{equation}
Thus, as $\gamma(z)$ varies from $0$ to $+\infty$, the angle $\alpha(z)$ varies from $\alpha(0)$ to $0$, or equivalently $\cos\alpha(z)$ varies from $\cos\alpha(0)$ to $1$. Fig. \ref{figmalus}b shows
$\alpha(z)$ as a function of $\alpha(0)$ for different values of the absorption term $\gamma(z)$.
Expressions (\ref{exp1}) and (\ref{exp2}) are dependent of the initial angle $\alpha(0)$. The integration of (\ref{exp1}) over all the possible initial angles $\alpha(0)$, \textit{i.e.}
\begin{equation}\label{coscos}
\left\langle\cos\alpha(z)\right\rangle_{\alpha(0)}=\displaystyle\frac 1\pi\int_0^\pi\,d\alpha(0)\,\cos\alpha(z),
\end{equation}
provides a convenient parameter to characterize the quality of the polarizer. 
Analytic calculation of (\ref{coscos}) gives us
\begin{equation}
\left\langle\cos\alpha(z)\right\rangle_{\alpha(0)}=\displaystyle\frac{1-\sqrt{1-\eta^2(z)}}{\eta(z)}=\tanh\displaystyle\frac{\gamma(z)}{2}.
\end{equation}
Hence, as $\gamma(z)$ varies from $0$ to $+\infty$ (perfect polarizer), the parameter $\left\langle\cos\alpha(z)\right\rangle_{\alpha(0)}$ varies from $0$ to $1$. 


\section{Composition law for polarizers}

From now on, we assume the $z$ parameter dependence of the different functions, such as \textit{e.g.} $\delta$ or $\gamma$, and
for clarity we drop the $z$ parameter from the following mathematical expressions.

\begin{figure}[t]
    \begin{center}
        \includegraphics[width=0.8\columnwidth]{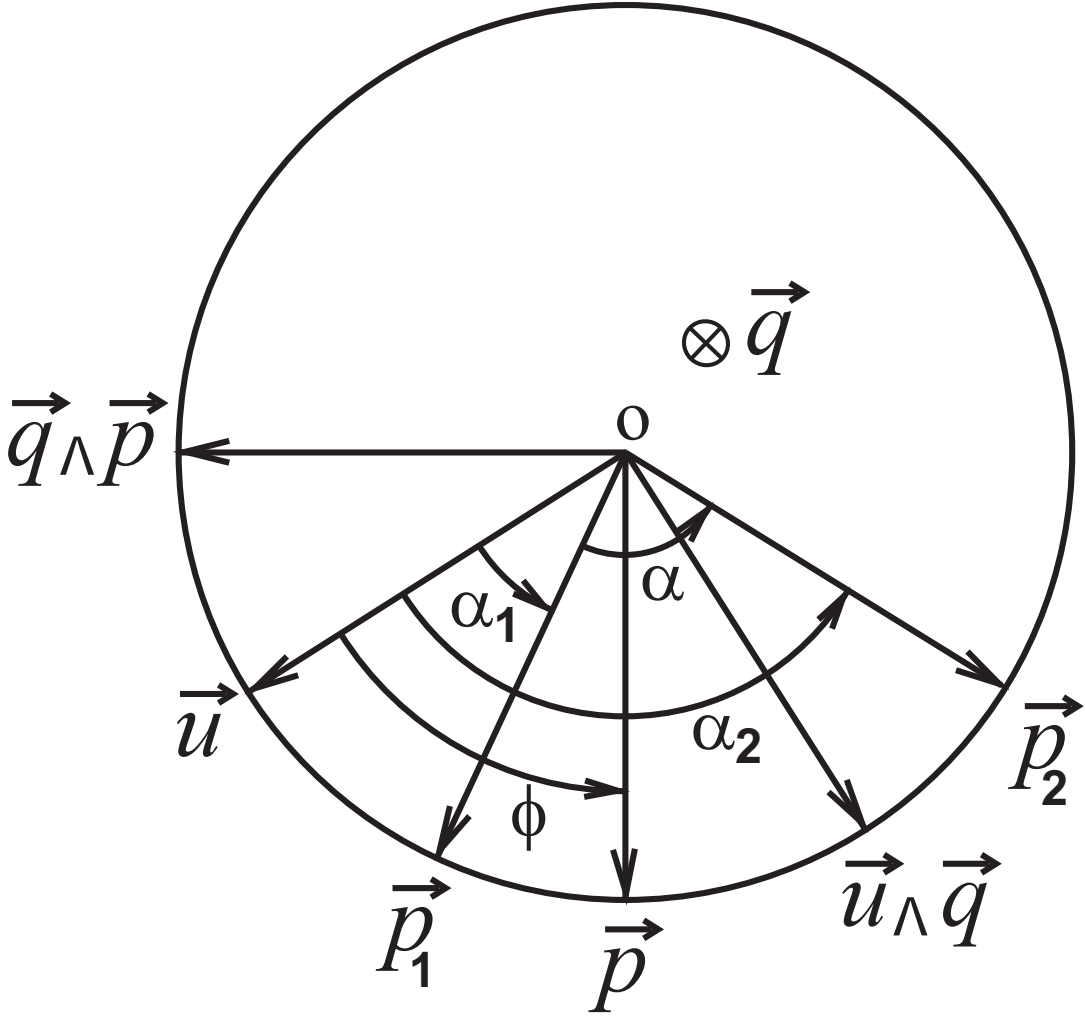}
    \end{center}
    \caption{\label{boost} Two successive boosts with respective direction $\vec p_1$ and $\vec p_2$
    can be decomposed in a boost of direction $\vec p$ and a rotation around the direction $\vec q$.}
\end{figure}

\subsection{Association of two polarizers}

Let us now consider the successive actions of two absorbing polarizers ($\delta=0$, see section \ref{sabs}) with different polarization axis $\ve p_1$ and $\ve p_2$, and
different absorption terms $\gamma_1$ and $\gamma_2$. The
corresponding operator of the total system is then
\begin{equation}\label{2boosts}
        P_{\vv{p_2},\gamma_2,0}P_{\vv{p_1},\gamma_1,0}=e^{-\left(\gamma_1+\gamma_2\right)/2}
        B_{\vv p_2,\gamma_2}
        B_{\vv p_1,\gamma_1}
\end{equation}
where we use the $SL(2,\mathbb{C})$ group representation (\ref{sl2cb}) of the Lorentz boost operator.
We keep in mind that the case of the composition of two general polarizers, \textit{i.e.} with $\delta\neq0$ (see section \ref{sec:gen}),
can be easily retrieved from the case (\ref{2boosts}) of two absorbing polarizers ($\delta=0$).
We focus now on the two successive boosts in (\ref{2boosts}). We know (see \textit{e.g.} \cite{Misner73,Halpern68})
that the action of two successive noncollinear
Lorentz boosts, $B_{\vv p_1,\gamma_1}$ and $B_{\vv p_2,\gamma_2}$, is equivalent to the action of a Lorentz boost $B_{\vv
p,\gamma}$ followed by a rotation $R_{\vv q,\delta}$, \textit{i.e.}
\begin{equation}\label{BBRB}
    B_{\vv p_2,\gamma_2}B_{\vv p_1,\gamma_1}=R_{\vv q,\delta}B_{\vv p,\gamma}
\end{equation}
with suitable parameters $\gamma$ and $\delta$, and suitable directions $\ve p$ and $\ve q$. Then, equation (\ref{2boosts}) can
be rewritten as
\begin{equation}\label{ppp1}
        P_{\vv{p_2},\gamma_2,0}P_{\vv{p_1},\gamma_1,0}=
        e^{-\left(\gamma_1+\gamma_2\right)/2}
        R_{\vv q,\delta}
        B_{\vv p,\gamma}
\end{equation}
or, using only polarizing device operators (\ref{op}), as
\begin{equation}\label{ppp2}
        P_{\vv{p_2},\gamma_2,0}P_{\vv{p_1},\gamma_1,0}=
        e^{\left(\gamma-\gamma_1-\gamma_2\right)/2}
        e^{i\delta/2}
        P_{\vv q,0,\delta}
        P_{\vv p,\gamma,0}.
\end{equation}

In the appendix, we derive from the initial parameters $\gamma_1$, $\gamma_2$, $\vv p_1$ and $\vv p_2$ the expressions of the resulting rotation parameters $\delta$ and $\vv q$ and of the resulting boost parameters $\gamma$ and $\vv p$ entering the relation (\ref{BBRB}) and consequently entering equations (\ref{ppp1}) and (\ref{ppp2}).
For the rotation parameters we have (\ref{qdef})
\begin{equation}\label{qdefb}
    \vv q=\frac{1}{\sin\alpha}\vv p_2\wedge\vv p_1
\end{equation}
and (\ref{delta})
\begin{equation}\label{ddef}
    \delta=2\arg\left(1+\tanh\displaystyle\frac{\gamma_1}{2}\tanh\displaystyle\frac{\gamma_2}{2}\,\,e^{i\alpha}\right)
\end{equation}
where $\alpha=\widehat{(\vv p_1,\vv p_2)}\in [0,\pi]$ is the angle between the two polarizer vectors $\ve p_1$
and $\ve p_2$. Equation (\ref{qdefb}) imply that the rotation axis $\vv q$ is orthogonal to the plane defined by the vectors
$\vv p_1$ and $\vv p_2$. Since the vectors $\vv p_1$, $\vv p_2$ and $\vv p$ belong to the same plane (see appendix section \ref{rotpar}),
it is possible to parametrize these vectors with the help of an arbitrary reference axis $\vv u$ belonging to this plane. Hence, the angles $\alpha_1=\widehat{(\vv u,\vv p_1)}$, $\alpha_2=\widehat{(\vv u,\vv p_2)}$, and $\phi=\widehat{(\vv u,\vv p)}$ characterize completely
the respective vectors $\vv p_1$, $\vv p_2$ and $\vv p$. Figure \ref{boost} gives an illustration of the relative positions of these vectors
and the associated angles. According to (\ref{eq5}), the boost parameters are given by the following expression
\begin{equation}\label{pgdef}
    \tanh\displaystyle\frac{\gamma}{2}\,\,e^{i\phi}=\displaystyle\frac{\tanh\displaystyle\frac{\gamma_1}{2}\,\,e^{i\alpha_1}+
    \tanh\displaystyle\frac{\gamma_2}{2}\,\,e^{i\alpha_2}}{1+\tanh\displaystyle\frac{\gamma_1}{2}\tanh\frac{\gamma_2}{2}
    \,\,e^{i\left(\alpha_2-\alpha_1\right)}}.
\end{equation}
Indeed, the modulus and the argument of (\ref{pgdef}) give respectively the absorption term $\gamma$ and the direction $\vv p$
of the resulting polarizer (through $\phi$ (\ref{phiphi})).

Let us now define the following \textit{addition} law
\begin{equation}\label{aoplusb}
    a\oplus b=\frac{a+b}{1+a^*b}, \qquad a,b\in\mathbb{C}.
\end{equation}
Using that law, equation (\ref{pgdef}) can be rewritten in the following more elegant form
\begin{equation}\label{voplusv}
    \Gamma e^{i\phi}=\Gamma_1 e^{i\alpha_1}\oplus \Gamma_2 e^{i\alpha_2}
\end{equation}
where we have have noted $\Gamma=\tanh\frac{\gamma}{2}$. Since
\begin{equation}
    P_{\vv{p_2},\gamma_2,0}P_{\vv{p_1},\gamma_1,0}\propto B_{\vv p_2,\gamma_2}B_{\vv p_1,\gamma_1}=R_{\vv q,\delta}B_{\vv p,\gamma}
\end{equation}
equation (\ref{voplusv}), which is similar to the addition law for non collinear velocities in special relativity
\cite{VigoureuxWigner}, is thus also the composition law for polarizers.

\subsection{Association of $N$ polarizers}

Consider now the action of $N$ successive absorbing polarizers (see section \ref{sabs}). The corresponding operator is
\begin{equation}\label{nb}
    \begin{array}{lcl}
        \mathcal{P}(1,2,\dots,N)
        &=&\overset{\curvearrowleft}{\prod}_{j=1}^N P_{\ve p_j,\gam_j,\zg}\\
        &=&P_{\ve p_N,\gam_N,\zg}\dots P_{\ve p_2,\gam_2,\zg}P_{\ve p_1,\gam_1,\zg}\\
        &=&\exp\left(-\frac 12\sum_{j=1}^N\gamma_j\right)\overset{\curvearrowleft}{\prod}_{j=1}^N B_{\ve p_j,\gam_j}\\
        &=&\exp\left(-\frac 12\sum_{j=1}^N\gamma_j\right) B_{\ve p_N,\gam_N}B_{\ve p_{N-1},\gam_{N-1}}
        \overset{\curvearrowleft}{\prod}_{j=1}^{N-2} B_{\ve p_j,\gam_j}\\
        &=&\exp\left(-\frac 12\sum_{j=1}^N\gamma_j\right) \mathcal{R}(N-1,N)\mathcal{B}(N-1,N)
        \overset{\curvearrowleft}{\prod}_{j=1}^{N-2} B_{\ve p_j,\gam_j}\\
        &=&\exp\left(-\frac 12\sum_{j=1}^N\gamma_j\right) \left(\overset{\curvearrowleft}{\prod}_{j=1}^{N-1}\mathcal{R}(j,\dots,N)\right)
        \mathcal{B}(1,\dots,N)
    \end{array}
\end{equation}
where we use the following notations
\begin{equation}
    B_{\ve p_j,\gam_j}B_{\ve p_i,\gam_i}=\mathcal{R}(i,j)\mathcal{B}(i,j)
\end{equation}
and
\begin{equation}
    \mathcal{B}(i,\dots,N)B_{\ve p_{i-1},\gam_{i-1}}=\mathcal{R}(i-1,\dots,N)\mathcal{B}(i-1,\dots,N).
\end{equation}
The operator $\mathcal{B}(i,\dots,N)$ is the boost operator resulting from the combination of the $(N-i)$ last boost operators.
The operator $\mathcal{R}(i,\dots,N)$ is the rotation operator resulting from the combination of the boost operators
$\mathcal{B}(i+1,\dots,N)$ and $B_{\ve p_i,\gam_i}$. For the sake of completeness, we define $\mathcal{B}(i)=B_{\ve p_i,\gam_i}$
and $\mathcal{R}(i)=\sigma^0$. In (\ref{nb}), the boost operator
\begin{equation}
    \mathcal{B}(1,\dots,N)=B_{\ve p,\gam}=e^{\frac{\gamma}{2}\ve p\cdot\ve \sigma}
\end{equation}
resulting from the combination of the $N$ boost operators $B_{\ve p_j,\gam_j}$ ($j=1,\dots,N$) is characterized by an absorption
parameter $\gamma$ and a vector $\ve p$ directly calculated from the $N-1$ successive applications of the composition law
(\ref{voplusv}). The ordered product in (\ref{nb})
\begin{equation}
    \overset{\curvearrowleft}{\prod}_{j=1}^{N-1}\mathcal{R}(j,\dots,N)=R_{\ve q,\del}=e^{\frac{i\delta}{2}\ve q\cdot\ve \sigma}
\end{equation}
is a resulting rotation operator characterized by a rotation vector $\ve q$ and an angle $\delta$ calculated from successive
applications of Eq. (\ref{delta}). The operator (\ref{nb}) corresponding to $N$ successive absorbing polarizers
can then be rewritten as
\begin{equation}\label{rbp}
    \mathcal{P}(1,2,\dots,N)=e^{-\frac 12\sum_{j=1}^N\gamma_j}R_{\ve q,\del}B_{\ve p,\gam}
\end{equation}
or using polarizing device operators (\ref{op}) as
\begin{equation}
    \mathcal{P}(1,2,\dots,N)=e^{\frac 12\left(\gamma-\sum_{j=1}^N\gamma_j\right)}e^{\frac i2 \delta}
    P_{\ve q,0,\del}P_{\ve p,\gam,0}.
\end{equation}

Using (\ref{rbp}), the initial state $\rho_{\ve{s}_0}$ of an incoming light wave is transformed into the following final state
\begin{equation}
    \rho_{\ve s}=\mathcal{P}(1,\dots,N)\rho_{\ve{s}_0}\mathcal{P}^\dagger(1,\dots,N)
\end{equation}
with the corresponding intensity
\begin{equation}
    I=\Tr{\left(\rho_{\ve s}\right)}=I_0e^{\gamma-\sum_{j=1}^N\gamma_j}\left(\cos^2\theta_0+\sin^2\theta_0e^{-2\gamma}\right).
\end{equation}
Here $2\theta_0$ (see section \ref{maluschap}) is the angle between the polarization $\ve s_0$ of the incoming wave and the
resulting polarization vector $\ve p$. So, the transmitted intensity of an unpolarized light beam is
\begin{equation}\label{malusNunp}
    \left\langle I\right\rangle_{\theta_0}
    =\frac{I_0}{2}e^{\gamma-\sum_{j=1}^N\gamma_j}\left(1+e^{-2\gamma}\right).
\end{equation}
Let us define $\eta_{\ve p}$ as the degree of polarization of the $N$ polarizers using the resulting polarization state
$\rho_{\ve p}$ as the state of reference. Using,
\begin{equation}
    I_{\ve p}=\Tr{\left(\rho_{\ve s}\rho_{\ve p}\right)}=I_0e^{\gamma-\sum_{j=1}^N\gamma_j}\cos^2\theta_0
\end{equation}
and
\begin{equation}
    I_{-\ve p}=\Tr{\left(\rho_{\ve s}\rho_{-\ve p}\right)}=I_0e^{-\gamma-\sum_{j=1}^N\gamma_j}\sin^2\theta_0
\end{equation}
we obtain the corresponding degree of polarization
\begin{equation}\label{ndop}
    \eta_{\ve p}=\frac{\left\langle I_{\ve p}\right\rangle_{\theta_0}-\left\langle I_{-\ve p}\right\rangle_{\theta_0}}
    {\left\langle I_{\ve p}\right\rangle_{\theta_0}+\left\langle I_{-\ve p}\right\rangle_{\theta_0}}=\tanh\gamma
\end{equation}
and the corresponding extinction ratio
\begin{equation}\label{nex}
    \xi_{\ve p}=\frac{\left\langle I_{-\ve p}\right\rangle_{\theta_0}}{\left\langle I_{\ve p}\right\rangle_{\theta_0}}=e^{-2\gamma}.
\end{equation}
The expressions found in (\ref{ndop}) and (\ref{nex}) are similar to the expressions of the degree of polarization $\eta$
(\ref{dop}) and of the extinction ratio $\xi$ (\ref{ex}) found for one polarizer except that here the absorption term $\gamma$
results from the action of $N$ absorbing polarizers. So, the composition law (\ref{voplusv}), whose the $N-1$ successive
application give $\Gamma=\tanh\frac\gamma2$, is also the composition law for the degree of polarization $\eta$ (\ref{ndop}).

To conclude this section let us retrieve the intensity of an initially unpolarized light beam through the succession of two
perfect polarizers ($\gamma_1\rightarrow\infty$, $\gamma_2\rightarrow\infty$). First, the modulus of the composition law (\ref{pgdef})
implies that $\gamma\rightarrow\infty$. In that case the expression (\ref{malusNunp}) becomes
\begin{equation}\label{malus2unp}
    \left\langle I\right\rangle_{\theta_0}
    =\frac{I_0}{2}e^{\gamma-\gamma_1-\gamma_2}.
\end{equation}
Using equations (\ref{eq1a}) and (\ref{tan}) in the case of two perfect polarizers ($\gamma_1\rightarrow\infty$, $\gamma_2\rightarrow\infty$, $\gamma\rightarrow\infty$) we obtain the following relation
\begin{equation}
    \cos^2\displaystyle\frac\alpha 2=e^{\gamma-\gamma_1-\gamma_2}
\end{equation}
where $\alpha=2\theta_{12}$ is the angle between the polarization vectors $\ve p_1$ and $\ve p_2$ (in the Poincar\'e sphere representation) of the two successive polarizers.
Thus Eq. (\ref{malus2unp}) can be rewritten as
\begin{equation}
    \left\langle I\right\rangle_{\theta_0}=\frac{I_0}{2}\cos^2\theta_{12}
\end{equation}
which is the well known expression for the intensity of an initially unpolarized light beam going through two perfect polarizers.
We remark that $\gamma$ given by the composition law (\ref{aoplusb}) encodes in (\ref{malus2unp}) and more generally in
(\ref{malusNunp}) the relative positions on the Poincar\'e sphere of the successive polarizer vectors $\ve p_j$.

\subsection{Discussion}

\begin{figure}
    \includegraphics[width=0.8\columnwidth]{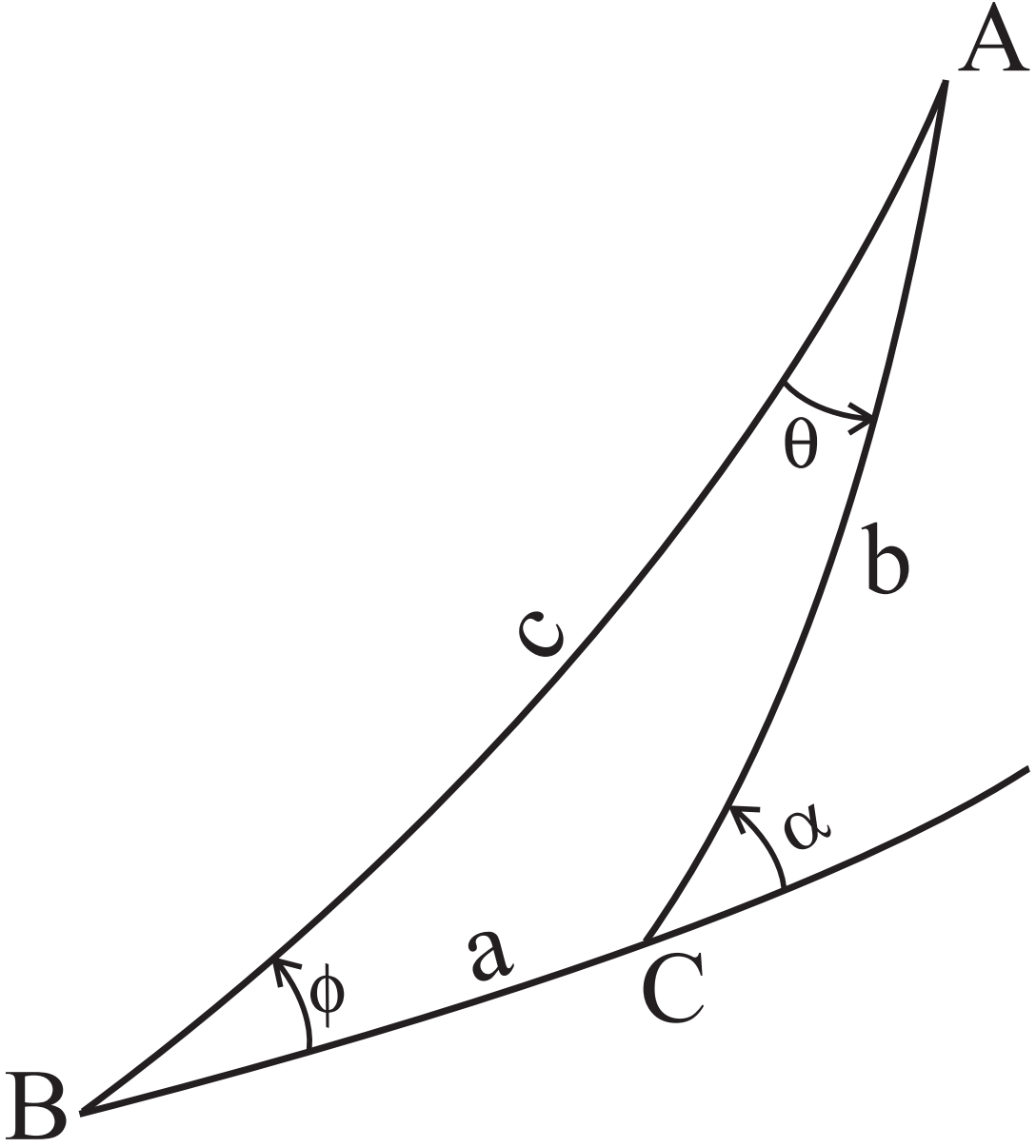}
    \caption{\label{triangle}The hyperbolic triangle ABC corresponding to the polarizers $P_1$, $P_2$ and to the resulting
    polarizer $P$ in the Poincar\'e disk.}
\end{figure}
The result (\ref{pgdef}) can be illustrated on the upper sheet $H^+$ of an hyperboloid, or, more easily, on its stereographic
projection known as the Poincar\'e disk. Let us consider in Fig. \ref{triangle} the hyperbolic triangle ABC, the three sides a,b, and c, of which corresponding to polarizers $P_1$, $P_2$ and to the resulting polarizer $P$ respectively. In that triangle, the
lengths of sides a,b, and c, opposite to vertices A,B, and C are  $\tanh\displaystyle\frac{\gamma_1}{2}$,
$\tanh\displaystyle\frac{\gamma_2}{2}$ and $\tanh\displaystyle\frac{\gamma}{2}$ respectively. They physically correspond to the
\textit{quality} of the corresponding polarizers: the length of each side approaches the limit $1$ when its corresponding
$\gamma$ tends to $\infty$. In that triangle (see Fig. \ref{triangle}), $\alpha$ is the angle between the axis of $P_1$ and the axis of $P_2$, $\phi$ is the angle between the axis of $P_1$ and the axis of the resulting polarizer $P$, and $\theta$ is the angle between the axis of $P$ and the axis of $P_2$. For the sake of simplicity, we choose the axis of reference along the direction of the first polarizers, \textit{i.e.} $\ve u=\ve p_1$, so that in Fig. \ref{boost} we have  $\alpha_1=0$ and $\alpha_2=\alpha$.

Now, in order to understand (\ref{pgdef}), let us use Fig. \ref{triangle} when the second polarizer $P_2$ is perfect, \textit{i.e.} $b=\tanh\displaystyle\frac{\gamma_2}{2}=1$. In such a case, (\ref{pgdef}) reduces to
\begin{equation}\label{aber1}
    \tanh\displaystyle\frac{\gamma}{2}\,e^{i\phi}=\displaystyle\frac{ \tanh\displaystyle\frac{\gamma_1}{2} +
    e^{i\alpha}}{1+\tanh\displaystyle\frac{\gamma_1}{2}\,e^{i\alpha}}.
\end{equation}
So, using (\ref{ddef}), we obtain
\begin{equation}\label{aber2}
    \tanh\displaystyle\frac{\gamma}{2}\,e^{i\phi}= e^{i\alpha}e^{-i\delta}.
\end{equation}
This result shows that the polarizer equivalent to using polarizers $P_1$ and $P_2$ successively (and in that order) is a perfect
polarizer (${\tanh\displaystyle\frac{\gamma}{2}=1}$) the axis of which is in the direction $\phi = \alpha-\delta$. That result
could be surprising, since we could expect to find its axis in the $\alpha$ direction. This comes from properties of
hyperbolic triangles. The angle $\theta$ between $P$ and $P_2$ can be calculated using (\ref{pgdef}): noting that the angle
of $AC$ with $CB$ is $-\alpha$ and taking now the direction of the arbitrary reference axis for angles in the direction of $P_2$,
we get (note that ${\tanh\displaystyle\frac{\gamma_2}{2}=1}$)
\begin{equation}\label{theta1}
    \tanh\displaystyle\frac{\gamma}{2} e^{-i\theta}= \frac{\tanh\displaystyle\frac{\gamma_2}{2}\,+
    \tanh\displaystyle\frac{\gamma_1}{2}\,\,e^{-i\alpha}}{1+\tanh\displaystyle\frac{\gamma_2}{2}\tanh\displaystyle\frac{\gamma_1}{2}\,\,e^{-i\alpha}}=
    \frac{1 + \tanh\displaystyle\frac{\gamma_1}{2}\,\,e^{-i\alpha}}{1+\tanh\displaystyle\frac{\gamma_1}{2}\,\,e^{-i\alpha}}=1
\end{equation}
so that $\theta=0$ : the axis of $P$ equivalent to using $P_1$ and $P_2$ successively is, as expected, directed in the direction of the axis of
$P_2$. These two last results, \textit{i.e.} the angle of $P$ with $P_1$ is $\phi = \alpha -\delta$ and that of $P$ with $P_2$ is $\theta = 0$, come
from the fact that the sum of the three angles of an hyperbolic triangle is not $\pi$, but $\pi -\delta$ where $\delta$ is its
angular defect. It is interesting to underline that this case ($P_2$ perfect) corresponds to the case of aberration in special relativity.

For the sake of completeness, let us show that $\delta$ is indeed the angular defect, that we note $\delta'$, of the hyperbolic triangle ABC.
The sum of the angles of the hyperbolic triangle ABC gives
\begin{equation}\label{sommeangle}
    {\phi + (\pi-\alpha) + \theta}= \pi-\delta'
\end{equation}
so that
    \begin{equation}\label{expsommeangle}
e^{i\delta'}= e^{i\alpha}e^{-i\phi}e^{-i\theta}.
\end{equation}
Using (\ref{phiphi}) and  extracting the angle $\theta$ from the first equality of (\ref{theta1}), Eq. (\ref{expsommeangle}) can be rewritten as
\begin{equation}\label{delta2}
    e^{i\delta'}= \frac{1 +\tanh\displaystyle\frac{\gamma_1}{2}\,\tanh\displaystyle\frac{\gamma_1}{2}\,\, e^{i\alpha}}{1
    +\tanh\displaystyle\frac{\gamma_1}{2}\,\tanh\displaystyle\frac{\gamma_1}{2}\,\, e^{-i\alpha}}.
\end{equation}
Comparing (\ref{delta2}) with (\ref{ddef}) gives us $\delta'=\delta$.
Thus the parameter $\delta$ of the rotation $R_{\vv q,\delta}$ (\ref{BBRB}) is the
angular defect of the hyperbolic triangle build using the parameters of $P_1$, $P_2$ and $P$.

To conclude this section, let us remark that the angle $\phi$ (\ref{phiphi}) giving the direction of the resulting polarizer $P$
can be rewritten as
\begin{equation}
\phi=\phi_{\mathrm{euclid}}-\frac\delta2
\end{equation}
where
\begin{equation}
\phi_{\mathrm{euclid}}=\arg\left(\tanh\frac{\gamma_1}{2}+\tanh\frac{\gamma_2}{2}e^{i\alpha}\right)
\end{equation}
would have been the angle between the axis of the polarizers $P_1$ and $P$ if the geometry were euclidean.

\section{Wigner angle}\label{chap:wigner}
\begin{figure}[t]
    \begin{center}
        \includegraphics[width=0.8\columnwidth]{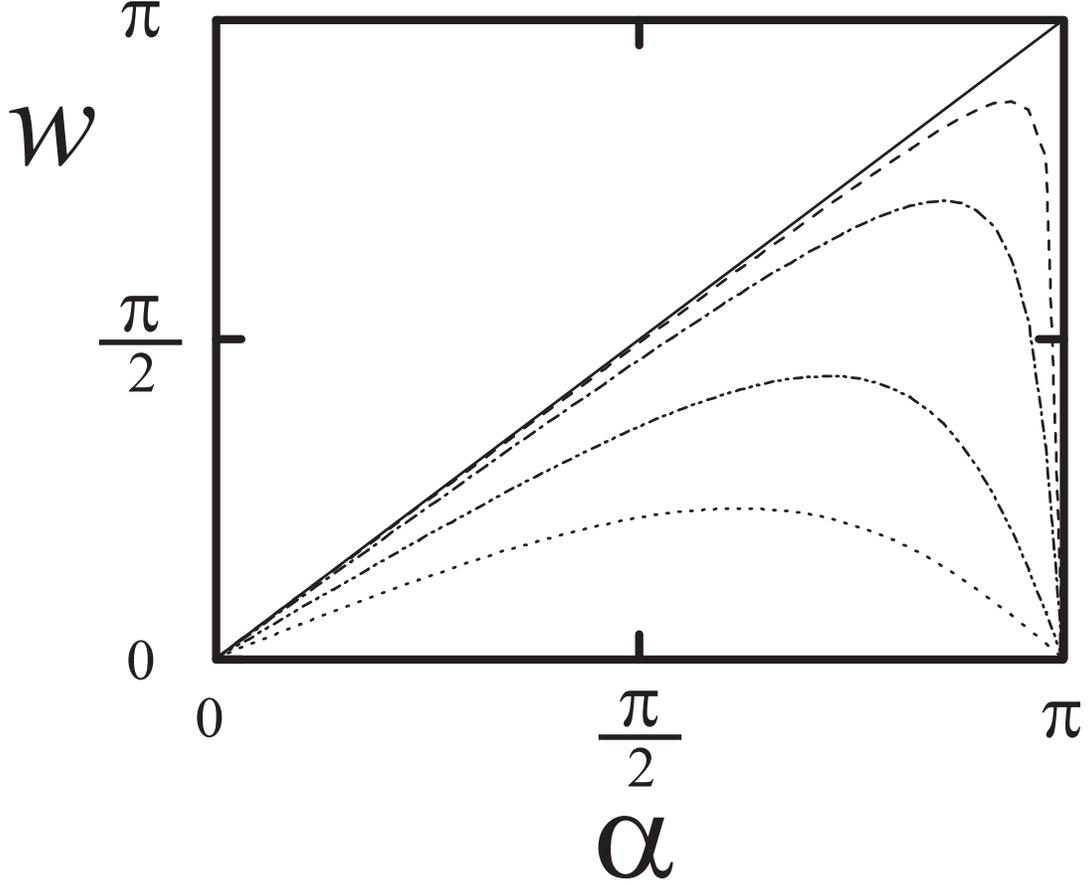}
    \end{center}
    \caption{\label{figwigner}Representation of the Wigner angle $w$ (\ref{wigner}) as a function of $\alpha$ for different value of
    the product $\sqrt{\Gamma_1\Gamma_2}$ from 0.6 (dot line), 0.8 (dot-dot-dash line), 0.95 (dot-dash line), 0.99 (dash line), 1 (solid line).}
\end{figure}
The composition law (\ref{voplusv}), \textit{i.e.} the addition law $\oplus$, is noncommutative \cite{Vigoureuxgroup}. Indeed, the final polarization of
the light wave passing through the polarizer 1 then through the polarizer 2 is not the same as the final polarization of the
light wave passing through the polarizer 2 then through the polarizer 1. Such a noncommutativity is obvious in the case of perfect
polarizers since the final polarization always corresponds to the polarization vector of the second polarizer. With two polarizers with
respective absorption terms $\gamma_1$ and $\gamma_2$ and with respective vectors $\ve p_1$ and $\ve p_2$, the two possibilities
of composition are
\begin{equation}\label{12}
    \Gamma_{_{1\oplus 2}} e^{i\phi_{1\oplus 2}}=\Gamma_1e^{i\alpha_1}\oplus \Gamma_2e^{i\alpha_2}
\end{equation}
and
\begin{equation}\label{21}
    \Gamma_{_{2\oplus 1}} e^{i\phi_{2\oplus 1}}=\Gamma_2e^{i\alpha_2}\oplus \Gamma_1e^{i\alpha_1}.
\end{equation}
From the definition of the addition law (\ref{aoplusb}) we easily see that
\begin{equation}
    \Gamma_{_{1\oplus 2}}=\Gamma_{_{2\oplus 1}}=\Gamma
\end{equation}
and consequently
\begin{equation}
    \gamma_{_{1\oplus 2}}=\gamma_{_{2\oplus 1}}=\gamma.
\end{equation}
However the angles $\phi_{1\oplus 2}$ and $\phi_{2\oplus 1}$ are not equal. Thus, by analogy with special relativity (or with optical studies of multilayers \cite{Vigoureux1}), we
introduce the Wigner angle $w=\phi_{2\oplus 1}-\phi_{1\oplus 2}$ which can be seen as a measure of the noncommutativity of the
addition law $\oplus$. Using (\ref{aoplusb}), (\ref{12}) and (\ref{21}), we easily determine the Wigner angle \cite{VigoureuxWigner} (see also \cite{Ungar1} and \cite{Ungar2}) through
\begin{equation}
    e^{iw}=\displaystyle\frac{\Gamma_2e^{i\alpha_2}\oplus \Gamma_1e^{i\alpha_1}}{\Gamma_1e^{i\alpha_1}\oplus \Gamma_2e^{i\alpha_2}}=\displaystyle\frac{1+\Gamma_1\Gamma_2e^{i\alpha}}{1+\Gamma_1\Gamma_2e^{-i\alpha}}=\Gamma_1\Gamma_2e^{i\alpha}\oplus1
\end{equation}
which gives then
\begin{equation}\label{wigner}
    w=2\arg\left(1+\Gamma_1\Gamma_2e^{i\alpha}\right)=\delta
\end{equation}
where $\alpha = \alpha_2 - \alpha_1$ is the angle between $\ve p_1$ and $\ve p_2$.

For perfect polarizers $\Gamma_1\Gamma_2=1$, the light wave is polarized first in one direction $\ve p_1$ then in the other
direction $\ve p_2$, so the Wigner angle is $w=\alpha$. For non perfect polarizers the measure of the Wigner angle $w$ can be seen as
a measure of the quality of the polarizers. Indeed, if one measures $w<\alpha$ for a given $\alpha$, we are able to retrieve the
product $\Gamma_1\Gamma_2$ and then to state on the quality of the polarizers. Fig. \ref{figwigner} shows the Wigner angle $w$ as
a function of $\alpha$ for different value of the product $\Gamma_1\Gamma_2$.

\section{Conclusion}
Textbooks usually only consider perfect polarizers (\textit{i.e.} polarizers with $\gamma\rightarrow\infty$) and do not examine the question of knowing how the polarization of a light beam is progressively transformed when propagating inside a polarizer. 
In the present paper, we have addressed these two problems using the notion of Stokes vector and that of the Poincar\'e sphere in order to characterize the evolution of a light wave polarization through a polarizer.
Whereas the action of a polarizer is intuitively thought of to be a simple rotation of the polarization along a geodesic on the Poincar\'e sphere, we have shown that it is not always the case.
In the general case, more complex trajectories differing from geodesics can in fact be found, especially in 
the case of total reflection based polarizers (see section \ref{sec:gen}). How these trajectories will affect geometric phases of light wave going through polarizers will be a question addressed in a following paper \cite{lgv2008}.

Non perfect polarizers, \textit{i.e.} here polarizers with finite $\gamma$, lead us naturally to define a generalized Malus law and a degree of polarization $\eta(z)$.

The action of a polarizer on light wave polarization can be described by Lorentz operators, where the attenuation factor $\gamma$ is the counterpart of the rapidity $\Phi$ commonly used in special relativity. The association of two such polarizers, involving then two Lorentz boosts, can always be viewed as the result of a Lorentz boost transformation followed by a rotation. In other words, the association of two polarizers is equivalent to the association of a polarizer and for example a device with optical activity giving an additional rotation of the polarization.
We have then derived a composition law (\ref{aoplusb}) for polarizers similar to the composition law for noncollinear velocities in special relativity.
This composition law $\oplus$ appears to be a very natural law to ``add'' quantities the values of which are limited to finite values. In fact, equation (\ref{aoplusb}) implies that no matter what are the values of the complex quantities $a$ and $b$, such as $\left|a\right| < 1$ and $\left|b\right| < 1$, the modulus of the overall resulting quantity $a\oplus b$ cannot exceed unity. Because it avoids infinity, such a generalization of Einstein's composition law of velocities appears to be a natural ``addition'' law of physical quantities in a closed interval. 

The association of $N$ polarizers can also be viewed as the association of a resulting polarizer and an optical device giving an additional rotation. Recursive iterations of the composition law (\ref{aoplusb}) give directly the characteristics of the resulting polarizer, \textit{i.e.} $\gamma$ and $\ve p$, and of its associate rotation, \textit{i.e.} $\delta$ and $\ve q$. We have defined the Malus law for the association of $N$ polarizers and the corresponding degree of polarization, the latter being easily determined using the composition law.

In addition to the degree of polarization (or equivalently the extinction ratio) we have defined others quantities characterizing the quality of the polarizers. We have defined, see section \ref{maluschap}, the angle $\alpha(z)$ measuring how far is the light wave polarization vector from the polarizer vector $\vv p$. Also, using the existing isomorphism between special relativity and polarization in optics, we have defined, see section \ref{chap:wigner}, the relativistic Wigner angle for the association of two polarizers. The Wigner angle is a direct consequence of the noncommutativity of the addition law $\oplus$ (\ref{aoplusb}) and we have shown that the latter can be used to directly measure the quality of the
association of two polarizers.

\appendix*

\section{}

We aim to determine the absorption term $\gamma$, the angle $\delta$, and the vectors $\ve p$ and $\ve q$ entering the relation
(\ref{BBRB}), \textit{i.e.}
\begin{equation}\label{BBRB1}
    B_{\vv p_2,\gamma_2}B_{\vv p_1,\gamma_1}=R_{\vv q,\delta}B_{\vv p,\gamma}.
\end{equation}
We define $\alpha=\widehat{(\vv p_1,\vv p_2)}\in [0,\pi]$ as the angle between the two polarizer vectors $\ve p_1$
and $\ve p_2$. Using the definitions (\ref{sl2cb}) and (\ref{sl2cr}) and expanding the exponential operators in (\ref{BBRB1}) we obtain the following set of  four equations
    \begin{equation}
        \label{eq1a}
        \cosh\displaystyle\frac{\gamma_2}{2}\cosh\displaystyle\frac{\gamma_1}{2}+
        \sinh\displaystyle\frac{\gamma_2}{2}\sinh\displaystyle\frac{\gamma_1}{2}
        \cos\alpha
        =
        \cos\displaystyle\frac{\delta}{2}\cosh\displaystyle\frac{\gamma}{2}
     \end{equation}
        
     \begin{equation}
        \label{eq1b}
        \sin\displaystyle\frac{\delta}{2}\sinh\displaystyle\frac{\gamma}{2}\vv
        q\cdot\vv p=0
     \end{equation}
        
     \begin{equation}
        \label{eq2a}
        \sinh\displaystyle\frac{\gamma_2}{2}\cosh\displaystyle\frac{\gamma_1}{2}\vv p_2
        +\sinh\displaystyle\frac{\gamma_1}{2}\cosh\displaystyle\frac{\gamma_2}{2}\vv p_1 =
        \sinh\displaystyle\frac{\gamma}{2}\cos\displaystyle\frac{\delta}{2}\vv p
        -\sin\displaystyle\frac{\delta}{2}\sinh\displaystyle\frac{\gamma}{2}\vv q\wedge \vv p
     \end{equation}

     \begin{equation}
        \label{eq2b}
        \sinh\displaystyle\frac{\gamma_2}{2}\sinh\displaystyle\frac{\gamma_1}{2}\vv
        p_2\wedge\vv p_1 =
        \sin\displaystyle\frac{\delta}{2}\cosh\displaystyle\frac{\gamma}{2}\vv q.
     \end{equation}

\subsection{Rotation parameters: $\delta$, $\vec q$}\label{rotpar}
As the equation (\ref{eq1b}) holds for any parameter $\gamma\in\mathbb{R}^+$ and for any angle $\delta\in[0,2\pi[$, we have $\vv
q\cdot\vv p=0$. Consequently, the two vectors $\ve p$ and $\ve q$ are orthogonal. Equation (\ref{eq2b}) fixes then the unit vector
$\ve q$ as
\begin{equation}\label{qdef}
    \vv q=\frac{1}{\sin\alpha}\vv p_2\wedge\vv p_1.
\end{equation}
The vectors $\vv p_1,\vv p_2$ and $\vv p$ are coplanar since $\ve p$ is orthogonal to $\ve q$ (see Fig. \ref{boost}). Using
(\ref{qdef}), equation (\ref{eq2b}) can be replaced by the following equation
\begin{equation}\label{eq2bb}
    \sinh\frac{\gamma_2}{2}\sinh\frac{\gamma_1}{2}\sin\alpha=\sin\frac{\delta}{2}\cosh\frac{\gamma}{2}.
\end{equation}
With equations (\ref{eq1a}) and (\ref{eq2bb}) the angle $\delta$ is given by
\begin{equation}\label{tan}
    \tan\displaystyle\frac \delta 2=\displaystyle\frac{\tanh\displaystyle\frac{\gamma_1}{2}
    \tanh\displaystyle\frac{\gamma_2}{2}\sin\alpha}{1+\tanh\displaystyle\frac{\gamma_1}{2}\tanh\displaystyle\frac{\gamma_2}{2}\cos\alpha}
\end{equation}
or equivalently by
\begin{equation}\label{delta}
    \displaystyle\frac \delta 2=\arg\left(1+\tanh\displaystyle\frac{\gamma_1}{2}\tanh\displaystyle\frac{\gamma_2}{2}\,\,e^{i\alpha}\right).
\end{equation}

Thus, from the initial parameters, $\gamma_1$, $\gamma_2$, $\ve p_1$ and $\ve p_2$, equations (\ref{qdef}) and (\ref{delta}) can be
used to obtain respectively the axis $\ve q$ and the angle $\delta$ of the rotation entering (\ref{BBRB1}).

\subsection{Boost parameters: $\gamma$, $\vec p$}
Let us now define the angles $\alpha_1=\widehat{(\vv u,\vv p_1)}$ and $\alpha_2=\widehat{(\vv u,\vv p_2)}$ (see Fig. \ref{boost})
where $\vv u$ is an arbitrary reference axis in the plane defined by $\vv p_1$ and $\vv p_2$. In order to calculate the angle
$\phi=\widehat{(\vv u,\vv p)}$, we project the two members of the equation (\ref{eq2a}) along the direction $\ve u$ and along the
orthogonal direction $\vv u \wedge \vv q$. We obtain the following two equations
\begin{widetext}
    \begin{eqnarray}
        \sinh\frac{\gamma_2}{2}\cosh\frac{\gamma_1}{2}\cos\alpha_2 +\sinh\frac{\gamma_1}{2}\cosh\frac{\gamma_2}{2}\cos\alpha_1
        =\cos\frac{\delta}{2}\sinh\frac{\gamma}{2}\cos\phi-\sin\frac{\delta}{2}\sinh\frac{\gamma}{2}\sin\phi\\
        \notag
        \\
        \sinh\frac{\gamma_2}{2}\cosh\frac{\gamma_1}{2}\sin\alpha_2 +\sinh\frac{\gamma_1}{2}\cosh\frac{\gamma_2}{2}\sin\alpha_1
        =\cos\frac{\delta}{2}\sinh\frac{\gamma}{2}\sin\phi+ \sin\frac{\delta}{2}\sinh\frac{\gamma}{2}\cos\phi.
    \end{eqnarray}
\end{widetext}
The combination of these two equations gives us
\begin{equation}\label{eq3}
    \sinh\frac{\gamma_2}{2}\cosh\frac{\gamma_1}{2}e^{i\alpha_2}+ \sinh\frac{\gamma_1}{2}\cosh\frac{\gamma_2}{2}e^{i\alpha_1}
    =\sinh\frac{\gamma}{2}e^{i\left(\phi+\delta/2\right)}.
\end{equation}
Combining now (\ref{eq1a}) and (\ref{eq2bb}), we obtain
\begin{equation}\label{eq4}
    \cosh\frac{\gamma_2}{2}\cosh\frac{\gamma_1}{2}+
    \sinh\frac{\gamma_2}{2}\sinh\frac{\gamma_1}{2} e^{i\alpha} =
    \cosh\frac{\gamma}{2}e^{i\delta/2}.
\end{equation}
Dividing the members of equation (\ref{eq3}) by those of equation (\ref{eq4}), we find
\begin{equation}\label{eq5}
    \tanh\displaystyle\frac{\gamma}{2}\,\,e^{i\phi}=\displaystyle\frac{\tanh\displaystyle\frac{\gamma_1}{2}\,\,e^{i\alpha_1}+
    \tanh\displaystyle\frac{\gamma_2}{2}\,\,e^{i\alpha_2}}{1+\tanh\displaystyle\frac{\gamma_1}{2}\tanh\frac{\gamma_2}{2}
    \,\,e^{i\left(\alpha_2-\alpha_1\right)}}.
\end{equation}

From the initial parameters, $\gamma_1$, $\gamma_2$, $\ve p_1$ and $\ve p_2$, equation (\ref{eq5}) allows us to calculate the
absorption term $\gamma$ and the direction $\ve p$ entering (\ref{BBRB1}). Indeed, the modulus of (\ref{eq5}) gives us $\gamma$
since
\begin{equation}\label{thgam}
    \tanh\displaystyle\frac{\gamma}{2}=\left|\displaystyle\frac{\tanh\displaystyle\frac{\gamma_1}{2}+
    \tanh\displaystyle\frac{\gamma_2}{2}\,\,e^{i\left(\alpha_2-\alpha_1\right)}}{1+\tanh\displaystyle\frac{\gamma_1}{2}\tanh\frac{\gamma_2}{2}
    \,\,e^{i\left(\alpha_2-\alpha_1\right)}}\right|
\end{equation}
and the direction of the unit vector $\ve p$ is determined from
\begin{equation}\label{phiphi}
    \phi=\alpha_1+\arg\left(\displaystyle\frac{\tanh\displaystyle\frac{\gamma_1}{2}+
    \tanh\displaystyle\frac{\gamma_2}{2}\,\,e^{i\left(\alpha_2-\alpha_1\right)}}
    {1+\tanh\displaystyle\frac{\gamma_1}{2}\tanh\displaystyle\frac{\gamma_2}{2}\,\,e^{i\left(\alpha_2-\alpha_1\right)}}\right).
\end{equation}

\end{document}